\documentclass[]{spie}  

\usepackage{amsmath,amsfonts,amssymb}
\usepackage{graphicx}
\usepackage[colorlinks=true, allcolors=blue]{hyperref}

\usepackage{graphicx}
\usepackage{subcaption}
\usepackage[export]{adjustbox}
\usepackage{wrapfig}

\usepackage{multirow}

\usepackage{float}

\title{Optical Design and Wavelength Calibration of a DMD-based Multi-Object Spectrograph}

\author[a]{Shaojie Chen}
\author[a,b]{Matthew C. H. Leung}
\author[c]{Xuefeng Yao}
\author[a,d]{Suresh Sivanandam}
\author[e]{Isabelle Sanders}
\author[d]{Rosalind Liang}
\affil[a]{Dunlap Institute for Astronomy and Astrophysics, University of Toronto, Toronto, ON, Canada}
\affil[b]{Division of Engineering Science, University of Toronto, Toronto, ON, Canada}
\affil[c]{Changchun Institute of Optics, Fine Mechanics and Physics, Chinese Academy of Sciences, Changchun, JL, China}
\affil[d]{Department of Astronomy and Astrophysics, University of Toronto, Toronto, ON, Canada}
\affil[e]{Department of Mechanical and Aerospace Engineering, University of California, Los Angeles (UCLA), Los Angeles, CA, USA}

\authorinfo{Further author information: (Send correspondence to S. Chen or M.C.H. Leung)\\S. Chen: E-mail: sj.chen@utoronto.ca\\M.C.H. Leung: E-mail: matthewchingho.leung@mail.utoronto.ca}

\begin{document} 
\maketitle

\begin{abstract}
The multi-object spectrograph (MOS) has been the benchmark for the current generation of astronomical spectrographs, valued for its ability to acquire the spectra of hundreds of objects simultaneously. In the last two decades, the digital micromirror device (DMD) has shown potential in becoming the central component of the MOS, being used as a programmable slit array. We have designed a seeing-limited DMD-based MOS covering a spectral range of 0.4 to 0.7 \textmu m, with a field of view (FOV) of $10.5^\prime \times 13.98^\prime$ and a spectral resolution of $R\sim1000$. This DMD-MOS employs all-spherical refractive optics, and a volume phase holographic (VPH) grism as the dispersive element for high throughput. In this paper, we present the optical design and optimization process of this DMD-MOS, as well as a preliminary wavelength calibration procedure for hyperspectral data reduction. Using simulated data of the DMD-MOS, a procedure was developed to measure hyperspectral imaging distortion and to construct pixel-to-wavelength mappings on the detector. An investigation into the relationships between DMD micromirrors and detector pixels was conducted. This DMD-MOS will be placed on a 0.5 m diameter telescope as an exploratory study for future DMD-based MOS systems.
\end{abstract}



\section{Introduction}
\label{sec:intro} 

Multi-object spectroscopy is an important technique that is widely used in major ground-based telescopes. A multi-object spectrograph (MOS) can acquire the spectra of hundreds of objects simultaneously, often in crowded fields. As a result, much observing time is saved, allowing for a more efficient use of telescopes. For these reasons, MOS systems have been used for various scientific programs, such as large surveys of stars and galaxies \cite{Smee2013, Kashino2013, LeFevre2005}. In order to select the objects of interest, current MOS systems use repositionable optical fibres (e.g. Subaru FMOS\cite{Kimura2010}), custom-made non-generalizable slit masks (e.g. Gemini GMOS\cite{Davies1997}), or a microelectromechanical systems (MEMS) component as a programmable slit mask. One MEMS component that has been investigated for use in MOS systems is the micro-shutter array (MSA), which was manufactured by NASA and has been successfully applied on the James Webb Space Telescope NIRSpec\cite{Bagnasco2007}. However, considering the cost and technical challenges of the MSA, another type of MEMS component shows greater potential for use as a programmable slit mask in MOS systems: the digital micromirror device (DMD).

A DMD is an array of small mirrors, called micromirrors, which could be individually programmed to tilt in two configurations. DMDs were first manufactured by Texas Instruments (TI), and were conventionally used in commercial projectors for fast light modulation. Its applications have expanded to manufacturing (e.g. semiconductor lithography), telecommunications, and displays\cite{Dudley2003}, leading to a decrease in cost.  In the last two decades, the use of DMD-based MOS systems has been explored in studies for several telescopes such as the Gemini Observatory (GMOX\cite{Barkhouser2016}), the Kitt Peak National Observatory (IRMOS\cite{MacKenty2006}), and the Galileo National Telescope (BATMAN\cite{Zamkotsian2014}). Studies have shown that DMD-based MOS systems offer more advantages in comparison to other types of MOS systems. Optical fibre MOS systems have greater complexity, and take a longer time to select objects since fibres need to be repositioned\cite{Travinsky2017}. In addition, in comparison to conventional machined slit masks or large cryogenic slit forming mechanisms, DMDs are more generalizable and use less time in terms of object selection, have a lower cost, and allow for smaller sized systems. Moreover, in comparison to MSAs, DMDs are readily available on the consumer market at a lower cost. The DMD’s optical performance has also been evaluated and characterized in several studies, and has shown to perform reliably in environments similar to space\cite{Robberto2009,Travinsky2017,Canonica2012}. Next generation telescopes, especially those with size constraints such as space-based telescopes, would benefit from employing DMDs in MOS systems as a programmable slit mask.

In this paper, we present the design of a DMD-based MOS, in which the DMD is used a programmable slit due to its array of individually programmable small mirrors. Since each micromirror can tilt in two configurations, the DMD-MOS has the capabilities to acquire spectra and images at the same time, through a spectrograph channel and an imaging channel. The parallel imaging channel provides real-time monitoring of the optimal slit alignment over the full field of view (FOV) and accurate photometry for absolute flux calibration, needed for long-term monitoring of transients. The spectrograph channel that we have designed covers a spectral range of 0.4-0.7 \textmu m, with a FOV of $10.50^\prime \times 13.98^\prime$ and a spectral resolution of R$\sim$1000 at 550 nm. The optical components used were entirely based on spherical lenses with high transmission materials. A volume phase holographic (VPH) grism was used as the dispersive element to enhance the throughput and to help minimize aberrations. We will present the optical design process of the DMD-MOS, and show how optical components were optimized to minimize aberrations and distortion across the entire FOV. After the optical design, a preliminary wavelength calibration procedure was developed for hyperspectral data reduction based on simulated data on the detector of the spectrograph channel. This simulated data was used to construct pixel-to-wavelength mappings on the detector and to measure hyperspectral imaging distortion. A preliminary investigation into the relationships between the DMD micromirrors, wavelength, and detector pixels were also conducted. This DMD-MOS was developed as part of an exploratory study for the next generation of ground-based telescopes.


\section{Overview}

\begin{figure}[H]
    \centering
    \includegraphics[width=0.5\textwidth]{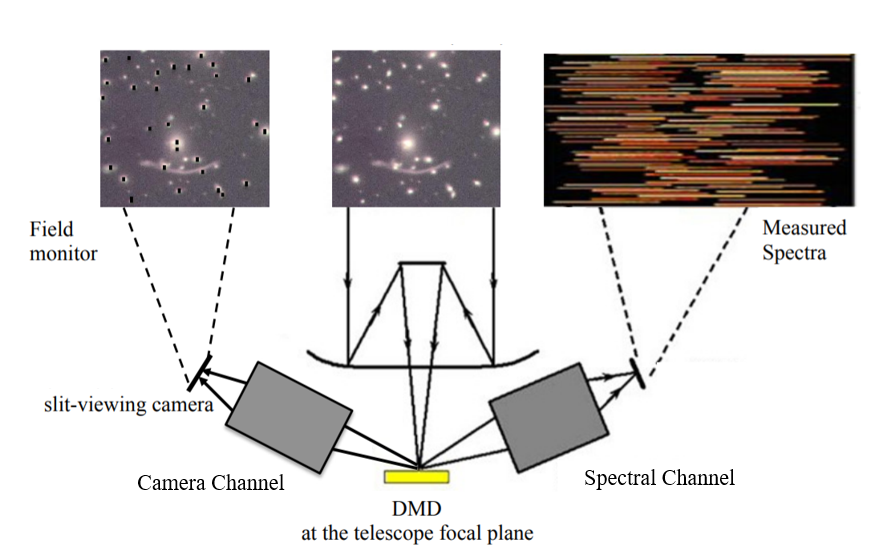}
    \caption{Double-channel of DMD-MOS, with the ON and OFF micromirrors feeding the spectrograph channel and imaging channel, respectively (Image Credit: SOAR Adaptive-Module Optical Spectrograph (SAMOS) Team\cite{Robberto2015,Smee2018})}
    \label{fig:double_channel_SAMOS}
\end{figure}

Our DMD-MOS is designed for seeing-limited operation with a CDK20 commercial telescope, which has a focal ratio of F/6.8 and a diameter of 0.5 m. The DMD-MOS has two channels: a spectrograph channel to acquire spectra of objects, and an imaging channel for standard imaging and object selection.  The structure of the double-channel DMD-MOS is shown in Figure \ref{fig:double_channel_SAMOS}.

Firstly, incoming light to the telescope is focused onto the DMD surface, which is located on the telescope focal plane. The DMD in our DMD-MOS is a TI DLP7000, which is an array of $768 \times 1024$ micromirrors each with a pitch of 13.7 \textmu m, sampling at $1.64^{\prime \prime}$ per micromirror. The DMD micromirrors can be individually programmed to tilt $\pm 12^\circ$, corresponding to an ON and an OFF configuration. Micromirrors in the OFF configuration reflect incoming light into the camera channel. Micromirrors in the ON configuration reflect incoming light into the spectrograph channel, acting as slits in a conventional spectrograph. Light reflected by the micromirrors into the spectrograph channel then pass through a collimator, and is then dispersed by a VPH grism. The dispersed light is then focused by some camera optics onto a detector. 

The DMD-MOS is able to resolve $1.64^{\prime \prime}$, which is the expected seeing resolution. It is designed to have a spectral resolution of at least $R\sim1000$ so that the redshifts of galaxies and atomic features in spectra could be adequately determined. In this paper, we will focus on the spectrograph channel design because of its complexity. For the imaging channel, which is mainly used for acquisition monitoring purposes, a traditional imaging design is adopted. A brief discussion of the imaging channel will be given. The parameters and requirements of the DMD-MOS are presented in Table \ref{table:1}. In addition to these requirements, we aimed to minimize the optical aberrations and distortion in the DMD-MOS, and to maximize throughput.

\begin{table}[h!]
\begin{center}
\begin{tabular}{ |l|l|l|} 
\hline
 & Parameter & Value \\
\hline
\hline
\multirow{2}{6em}{Telescope} 
& Diameter & 500 mm \\ 
& Output beam focal ratio & F/6.8 \\
\hline
\multirow{2}{6em}{DMD} 
& Micromirrors in use (X $\times$ Y direction) & $750\times1000$ \\
& Pitch of each micromirror & 13.7 \textmu m \\
\hline
\multirow{2}{10em}{Detector of Spectrograph Channel} 
& Pixels in use (X $\times$ Y direction) & 2000 $\times$ 2000 \\
& Pitch of each pixel & 6.5 \textmu m \\
\hline
\multirow{2}{10em}{Detector of Imaging Channel} 
& Pixels in use (X $\times$ Y direction) & 2046 $\times$ 1544 \\
& Pitch of each pixel & 3.45 \textmu m \\
\hline
\multirow{9}{10em}{Requirements of Spectrograph Channel} 
& Field of View & $10.50^\prime \times 13.98^\prime$ \\
& Plate Scale on DMD & $0.82^{\prime\prime}$ \\
& F/\# of Spectrograph &	F/3.3 \\
& Operating Wavelength Range & 0.40 \textmu m to 0.70 \textmu m \\
& Central Wavelength & 0.55 \textmu m \\
& Spectral Resolution & $\sim$1000 \\
& Seeing Resolution & $1.64^{\prime\prime}$ \\
& System Efficiency	& $>30\%$ \\
& Image Quality & EE50 @ $2\times2$ pixels \\
\hline
\multirow{4}{10em}{Requirements of Imaging Channel} 
& F/\# of Imager & F/3.4 \\
& Plate Scale on Detector & $0.42^{\prime\prime}$ \\
& System Efficiency	& $>50\%$ \\
& Imaging Quality & EE50 @ $4\times4$ pixels \\
\hline

\end{tabular}
\caption{Parameters and requirements}
\label{table:1}
\end{center}
\end{table}


\section{Design of DMD-MOS}

\subsection{Design Overview}

\begin{figure}[H]
    \centering
    \includegraphics[width=\textwidth]{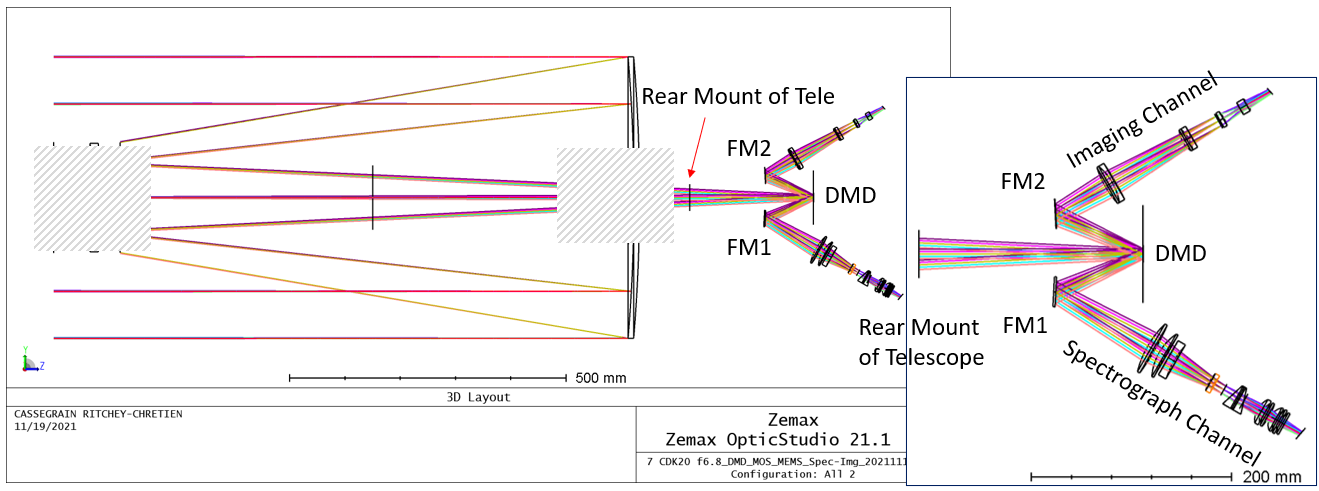}
    \caption{Overall layout of DMD-MOS; the front optics are not included because of a NDA}
    \label{fig:overall_layout}
\end{figure}

Our DMD-MOS is a lab prototype which is supported by seed funding, and hence its design was mainly limited by financial constraints and the input telescope optics. The current DMD-MOS is designed to be fed by a CDK20 commercial telescope. The overall optical layout is shown in Figure \ref{fig:overall_layout}. Note that a part of the front optics is not shown because of a nondisclosure agreement (NDA). The telescope is almost like a telecentric system, delivering an F/6.8 beam onto the focal plane, with small chief ray angles. The small angles of chief rays on the focal plane are ideal for increasing the efficiency of the DMD and for minimizing the crosstalk between adjacent micromirrors. Hence, the system is designed so that the DMD is located directly on the focal plane of the telescope. No relay optics between the telescope and the spectrograph are adopted. 

To provide parallel spectroscopic and acquisition modes, two symmetric channels are designed about the DMD. Since the DMD micromirrors can be individually programmed to tilt $\pm 12^\circ$, beams reflected by the micromirrors have angles $\pm 24^\circ$ with respect to the incident beam. The reflected beams then pass into the spectrograph channel and imaging channel. After reflection from the DMD, two flat fold mirrors (FM1 and FM2 in Figure \ref{fig:overall_layout}) are used to fold the optical path back forward, away from the rear of the telescope. However, limited by the large housing mount of the DMD itself, a large angle is required between the two channels. Since the DMD directs the beams off-axis, the deep angle of about $12^\circ$ on the image plane of DMD-MOS is implemented in order to compensate the tilt of the DMD. The large tilt angle on the detector drives the large back focal length, which is challenging for the optical design especially for the spectrograph channel due to the large spectra map across the full FOV. Both channels use spherical refractive lenses after the fold mirrors. The spectrograph and imaging channels generate spectral and geometric images respectively on the detectors, both with $\sim$0.5 magnification. These channels have focal ratios of F/3.3 and F/3.4 respectively. The FOV of the DMD-MOS is $10.50^\prime \times 13.98^\prime$, based on the resolution of $1.64^{\prime\prime}$ per micromirror on the DMD.

\begin{figure}[H]
    \centering
    \includegraphics[width=0.95\textwidth]{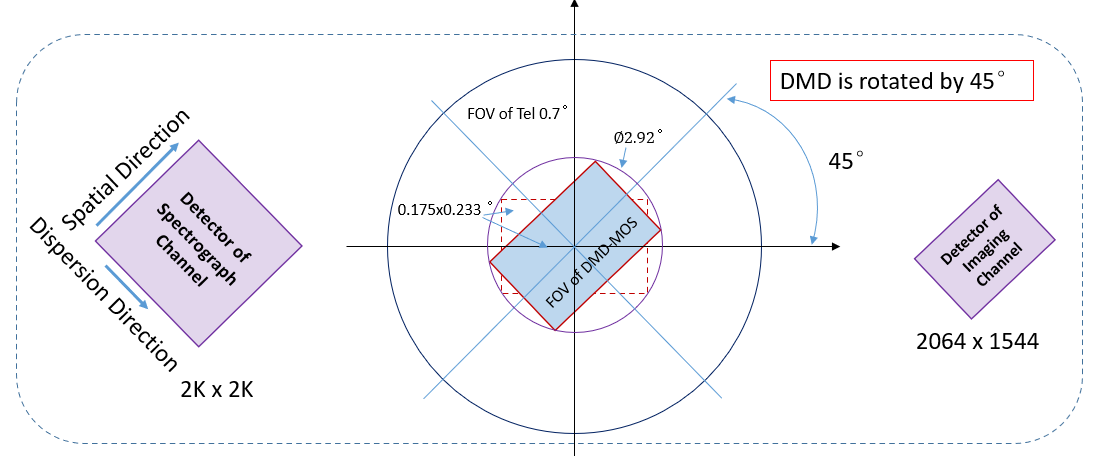}
    \caption{Orientations and relative relationship between the DMD and detectors}
    \label{fig:orientations}
\end{figure}

Due to the tile orientation of the micromirrors of DMD, the packaged DMD will rotated by $45^\circ$ about the optical axis. The detectors and the grism will be rotated by $45^\circ$ with respect to the DMD. The relative relationships between the FOV, DMD, and the detectors are shown in Figure \ref{fig:orientations}.

\begin{figure}[H]
    \centering
    \begin{subfigure}{0.49\textwidth}
        \centering
        \includegraphics[width=0.8\textwidth]{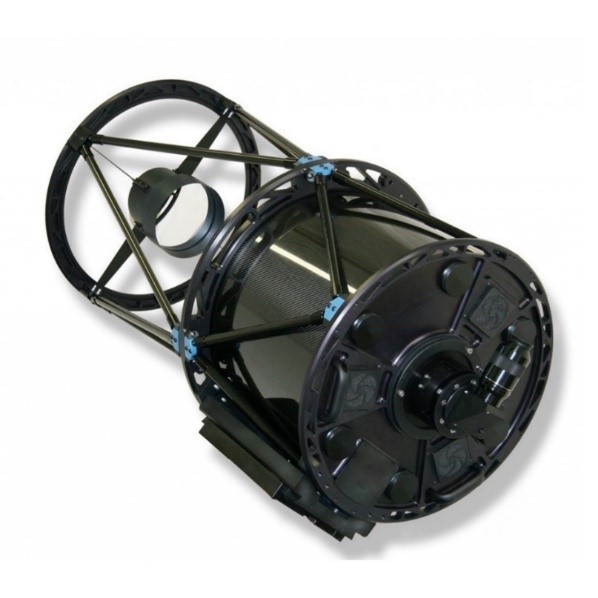}
        \caption{CDK20 Telescope}
    \end{subfigure}
    \begin{subfigure}{0.49\textwidth}
        \centering
        \includegraphics[width=0.8\textwidth]{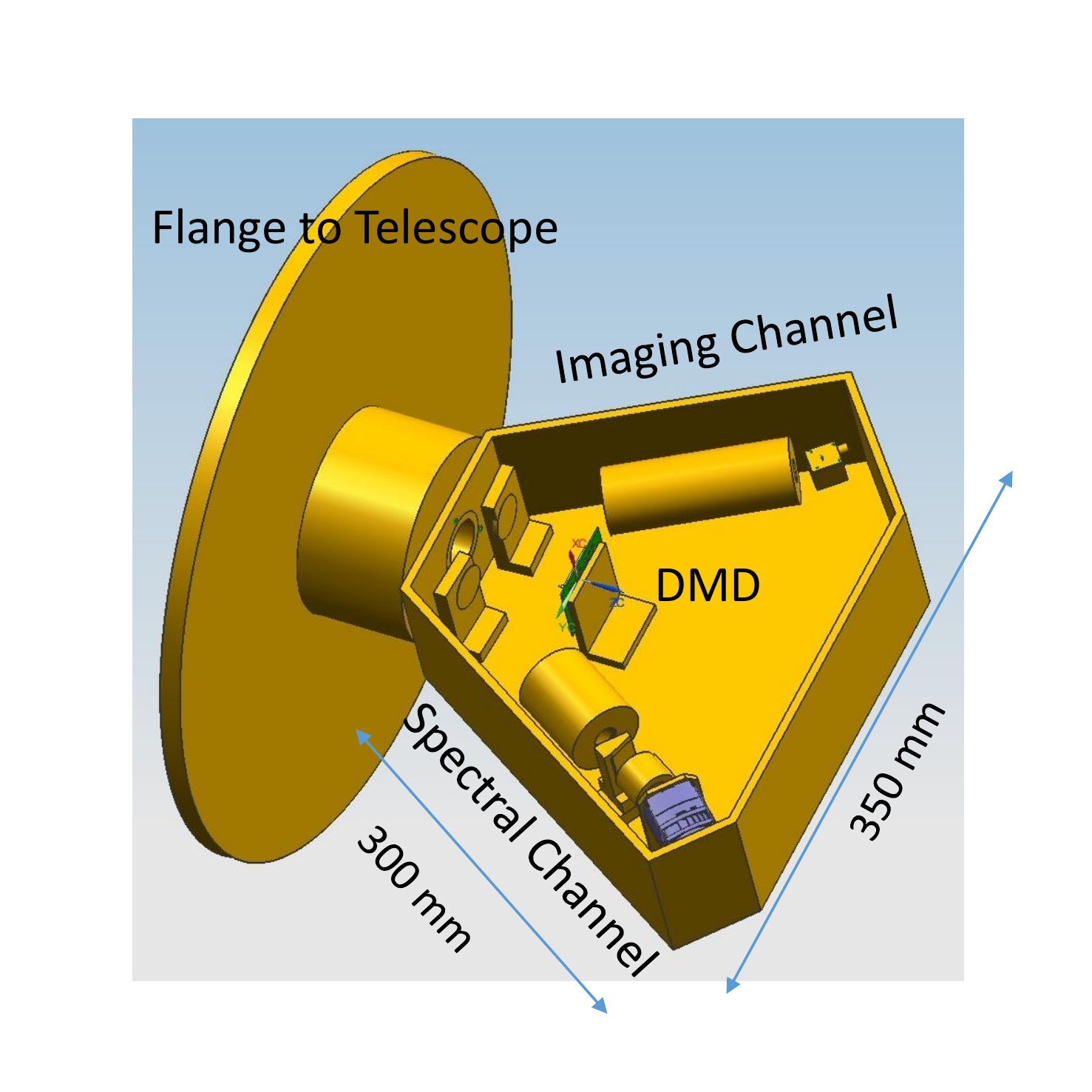}
        \caption{Opto-mechanical design of DMD-MOS}
    \end{subfigure}
    \caption{Telescope in front of the DMD-MOS and opto-mechanical design of the DMD-MOS}
    \label{fig:telescope_opto-mech}
\end{figure}

The overall opto-mechanical concept design of the DMD-MOS is shown in Figure \ref{fig:telescope_opto-mech}. The DMD-MOS will be attached to the back of the CDK20 telescope with a flange. The maximum load capacity of the telescope is 40 lb, and the total mass and the center of mass of the DMD-MOS has been analyzed to meet the requirements of the telescope.


\subsection{Key Components of the DMD-MOS}

\subsubsection{DMD Testing}

\begin{figure}[H]
    \centering
    \begin{subfigure}{0.49\textwidth}
        \centering
        \includegraphics[width=0.8\textwidth]{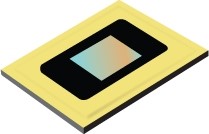}
        \caption{Packaged DMD DLP7000, from TI website}
    \end{subfigure}
    \begin{subfigure}{0.49\textwidth}
        \centering
        \includegraphics[width=0.8\textwidth]{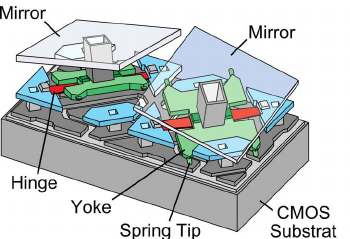}
        \caption{Two DMD pixels, mirrors can tilt perpendicular to the hinge axis (red); this graphic was from Ref. \citenum{Bayer2015} and Ref. \citenum{Douglass1998}.}
    \end{subfigure}
    \caption{Packaged DMD and typical substructure of a DMD}
    \label{fig:DMD}
\end{figure}

The DMD is the critical component of the DMD-MOS, and so the DMD’s features and capabilities need to be considered for the system design and for performance evaluation. Hence, lab testing of the DMD was conducted in order to evaluate several criteria, including scattered light, contrast performance, reflectance, and functional reliability. To test the DMD, a Gaussian laser beam with varying F/\#s was focused on the DMD, and a high-resolution micro imaging system was used for detection. From the testing, we found the scattered light level is below $1\%$ at 5 micromirrors away. But limited by the sensitivity of the detector, the scattered light was difficult to map adequately. From the testing, it was confirmed that most of the scatter was caused by the edges of the individual micromirrors. Using an extended light source, the reflectance efficiency of the DLP7000 was measured to be $(71\pm2)\%$. Our testing results agree with previous measurements carried out by other groups\cite{Travinsky2017}. The experiments show that loss mainly happens on the gaps of micromirrors. The typical packaged DMD and substructure of a TI DMD is shown in Figure \ref{fig:DMD}. 

\subsubsection{VPH Grism Design}

\begin{figure}[H]
    \centering
    \includegraphics[width=\textwidth]{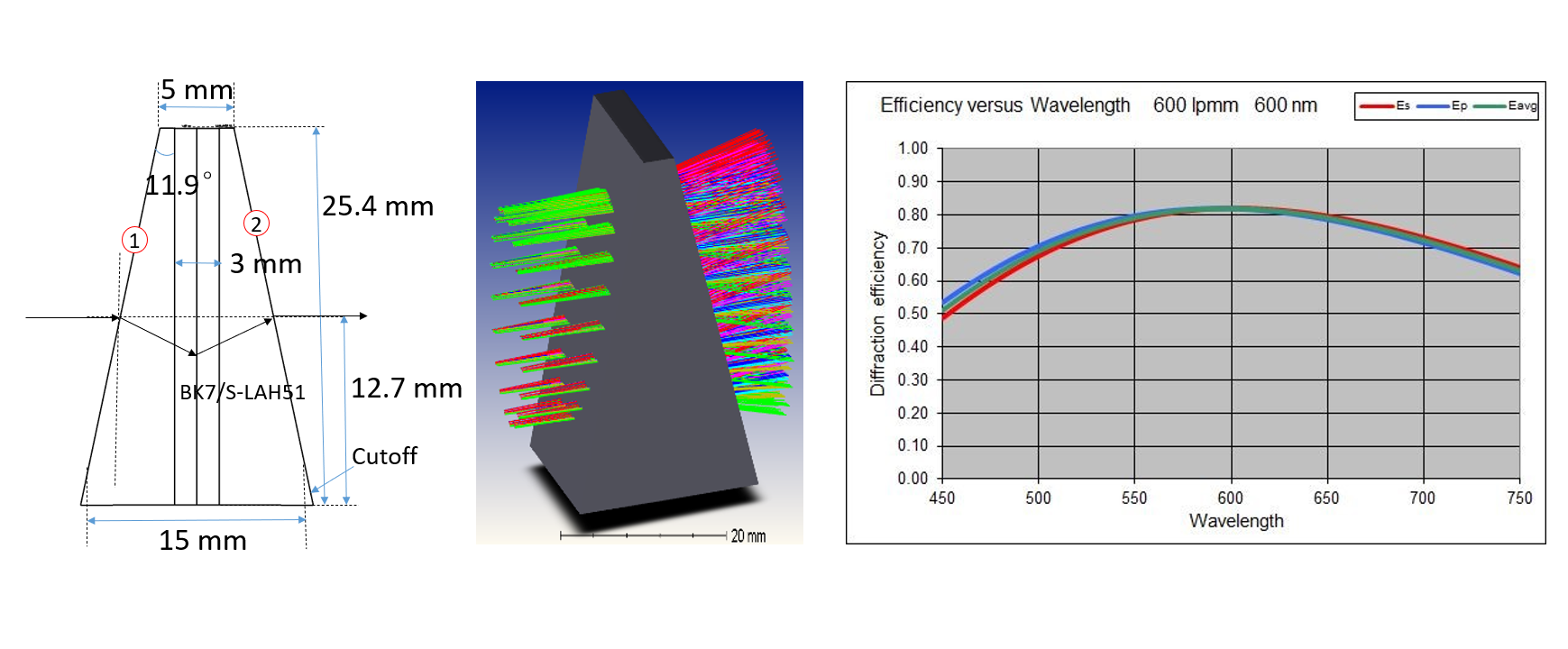}
    \caption{Design of VPH grism, and efficiency of VPH grating with 600 lines/mm}
    \label{fig:VPH}
\end{figure}

In the spectrograph channel, a VPH grism is adopted in order to achieve high throughput, and also to help optimize aberrations because of the consistent magnification of about 1.7 over the whole spectral band. The VPH grism consists of a VPH grating combined with two prisms\cite{Ebizuka2011,Arns2010,Nakajima2008}. In the VPH grism, the light needs to satisfy the Bragg condition as it passes through the volume of the film layer, shown as Figure \ref{fig:VPH}. The specific refractive index and prism apex angle are designed to compensate for the deviation caused by grating. The size of the grism is controlled at the same time. The VPH grism is designed so that there would be high transmission in the prisms and in the grating substrate across the full spectral band. Our VPH grism was custom-made from the vendor Wasatch Photonics. 

\subsubsection{Detector of Spectrograph Channel}

\begin{figure}[H]
    \centering
    \begin{subfigure}{0.49\textwidth}
        \centering
        \includegraphics[width=0.6\textwidth]{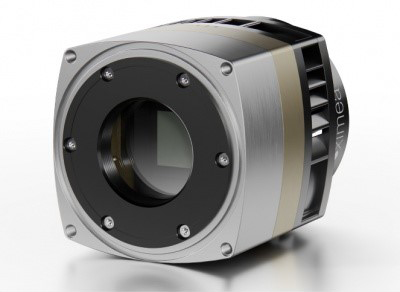}
        \caption{GSENSE2020BSI Detector}
    \end{subfigure}
    \begin{subfigure}{0.49\textwidth}
        \centering
        \includegraphics[width=0.66\textwidth]{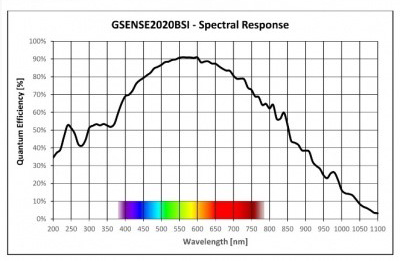}
        \caption{Detector quantum efficiency}
    \end{subfigure}
    \caption{Detector for the spectrograph channel and quantum efficiency of the detector}
    \label{fig:Detector}
\end{figure}

In order to increase the system sensitivity, a cooled backside-illuminated detector was selected for the spectrograph channel. Backside-illuminated detectors offer high spectral response and low readout noise. However, considering the cost of these detectors, a 2K by 2K detector was selected instead of one with a higher resolution. The detector in the spectrograph channel is a Gpixel GSENSE2020BSI CMOS detector. The quantum efficiency of this detector is shown in Figure \ref{fig:Detector}. 

\subsection{Optical Design and Performance of DMD-MOS}

\subsubsection{Performance of Spectrograph Channel}

\begin{figure}[H]
    \centering
    \includegraphics[width=0.8\textwidth]{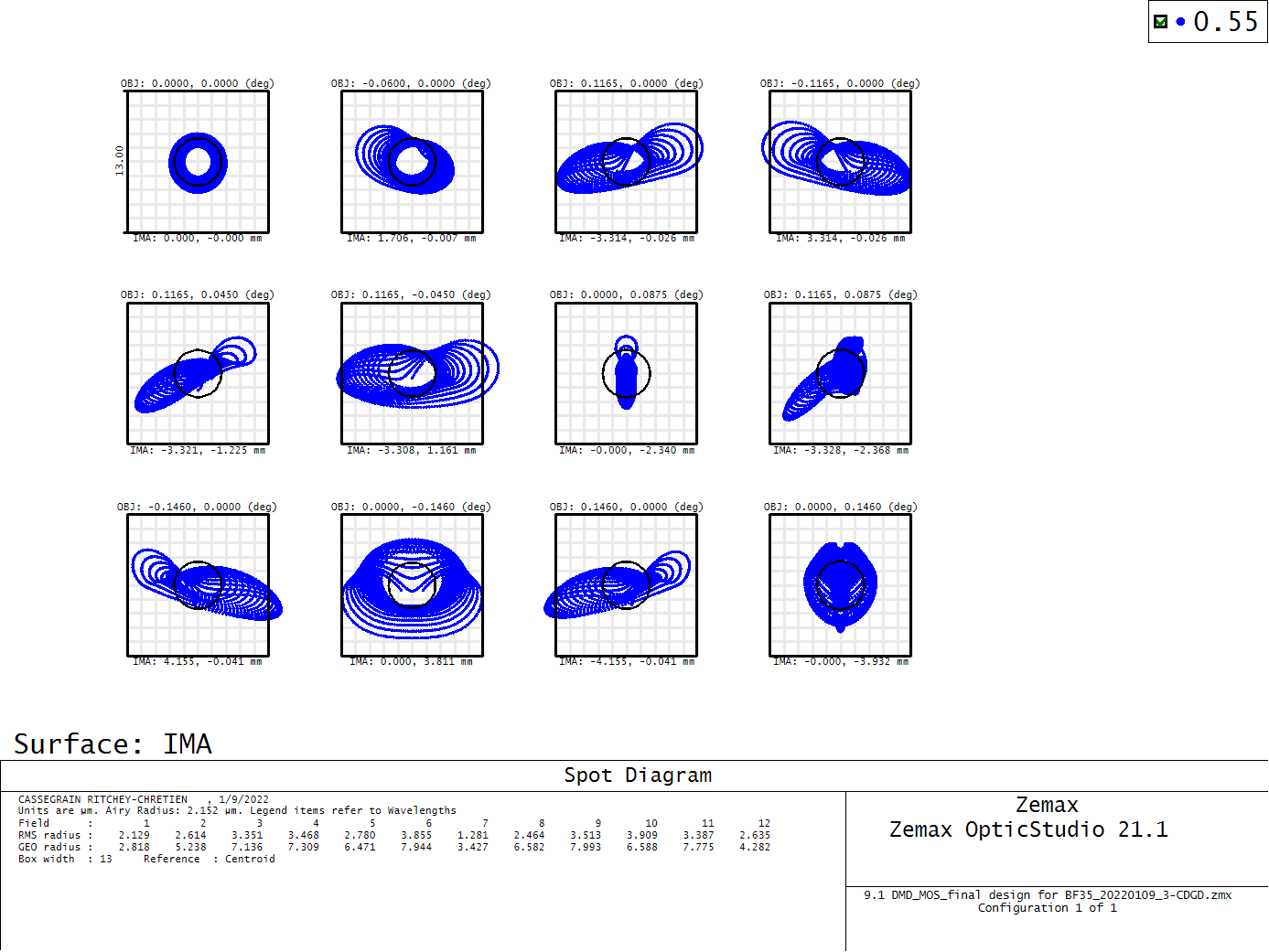}
    \caption{Spot diagrams at center wavelength 550 nm; each square is $2 \times 2$ pixels}
    \label{fig:550nm_spot}
\end{figure}

\begin{figure}[H]
    \centering
    \includegraphics[width=0.5\textwidth]{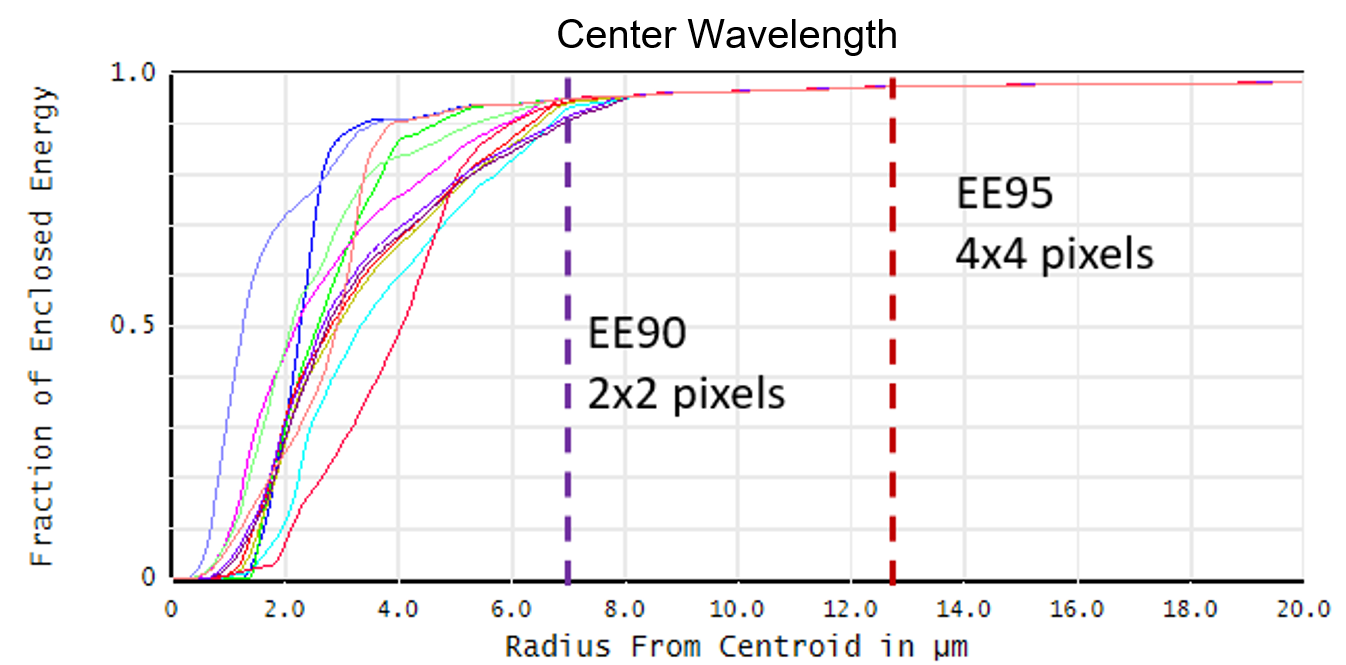}
    \caption{Enclosed energy curve for center wavelength 550 nm}
    \label{fig:550nm_EE}
\end{figure}

In the full FOV of the spectrograph ($10.5^\prime \times 13.98^\prime$), the whole spectral band is optimized by evaluating spot diagrams and enclosed energy. Due to the wide dispersion field, it is very challenging to achieve consistent performance across the entire band. The spot diagrams for the center wavelength (550 nm) across the full FOV are shown in Figure \ref{fig:550nm_spot}, and the enclosed energy curve is shown in Figure \ref{fig:550nm_EE}. The center wavelength has a great performance, which can achieve 90\% enclosed energy (EE90) within 2 by 2 pixels. The same evaluation method was applied to the edge wavelengths at 450 nm and 700 nm, and their performance degrades to some degree for certain fields, depending on the relationship between the FOV and spectrum location. However, the overall performance can meet the requirements of 50\% encircled energy (EE50) being within 2 by 2 pixels. Limited by the effective wavelength range of the telescope is 450 to 700 nm, we have to do a tradeoff between 400 nm and 450 nm. Thus, the spectrograph is designed from 400 to 700 nm, but the performance of 400 to 450 nm is significantly affected by the aberration of the telescope. The EE50 at 400 nm is around 4 by 4 pixels.

\begin{figure}[H]
    \centering
    \begin{subfigure}{0.39\textwidth}
        \centering
        \includegraphics[width=\textwidth]{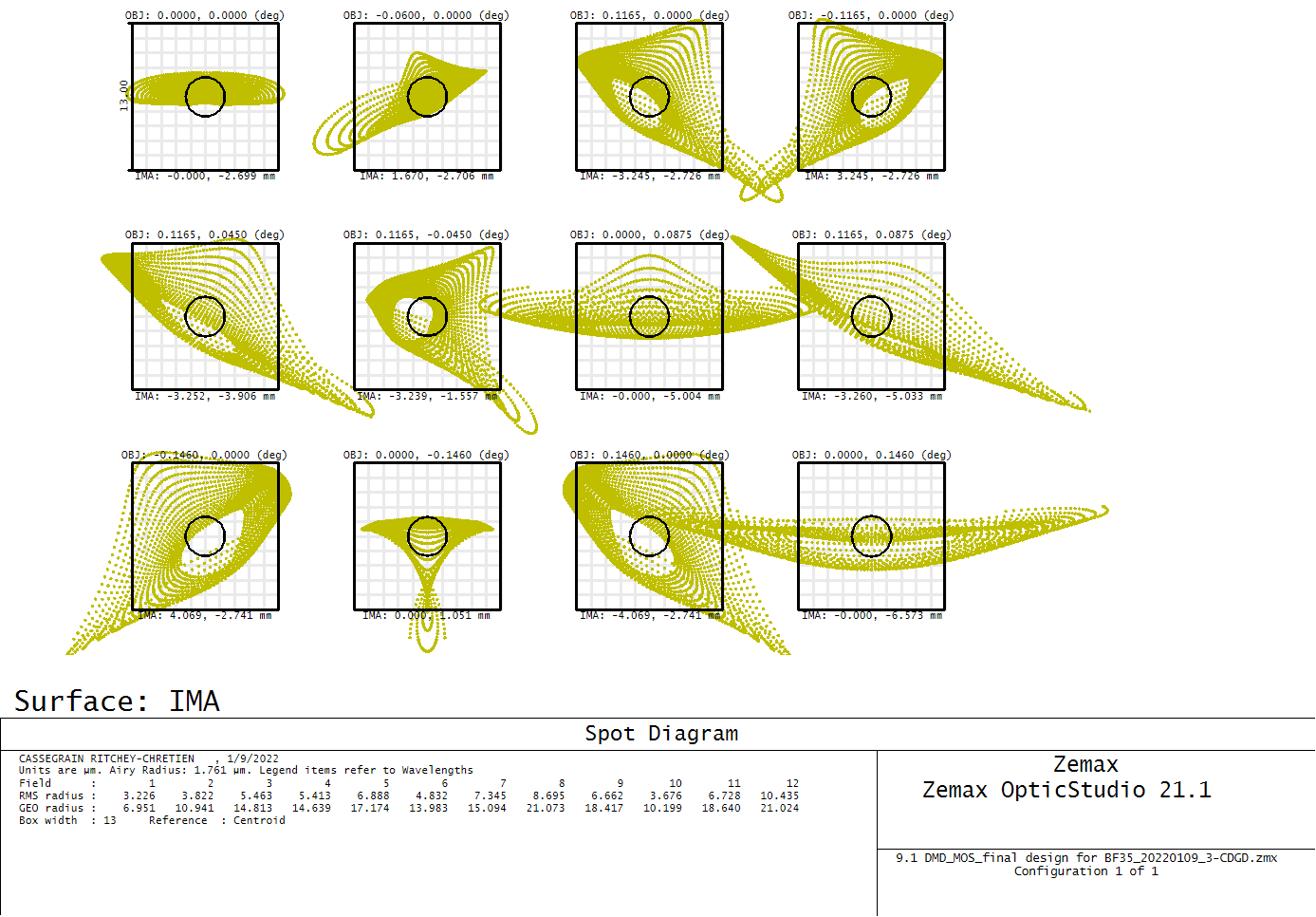}
        \caption{Spot diagrams at 450 nm }
    \end{subfigure}
    \begin{subfigure}{0.59\textwidth}
        \centering
        \includegraphics[width=0.9\textwidth]{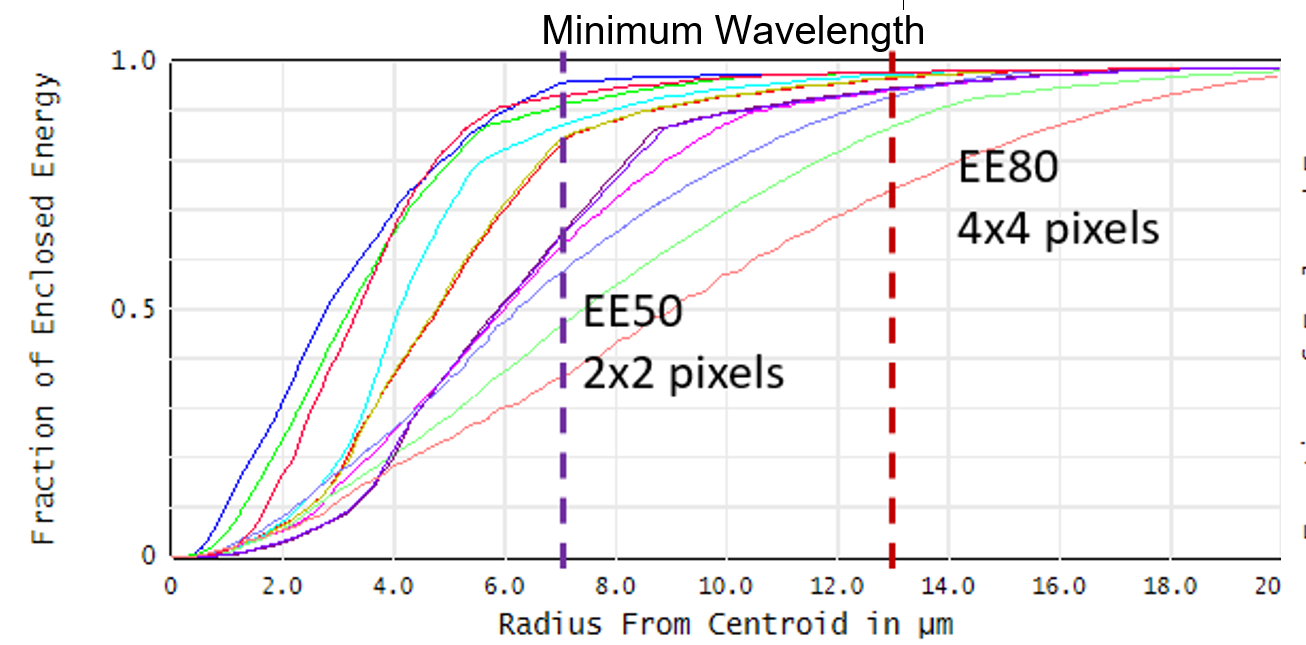}
        \caption{Enclosed energy curve at 450 nm}
    \end{subfigure}
    \caption{Performance for 450 nm; in the spot diagrams, each square is $2 \times 2$ pixels}
    \label{fig:450nm}
\end{figure}

\begin{figure}[H]
    \centering
    \begin{subfigure}{0.39\textwidth}
        \centering
        \includegraphics[width=\textwidth]{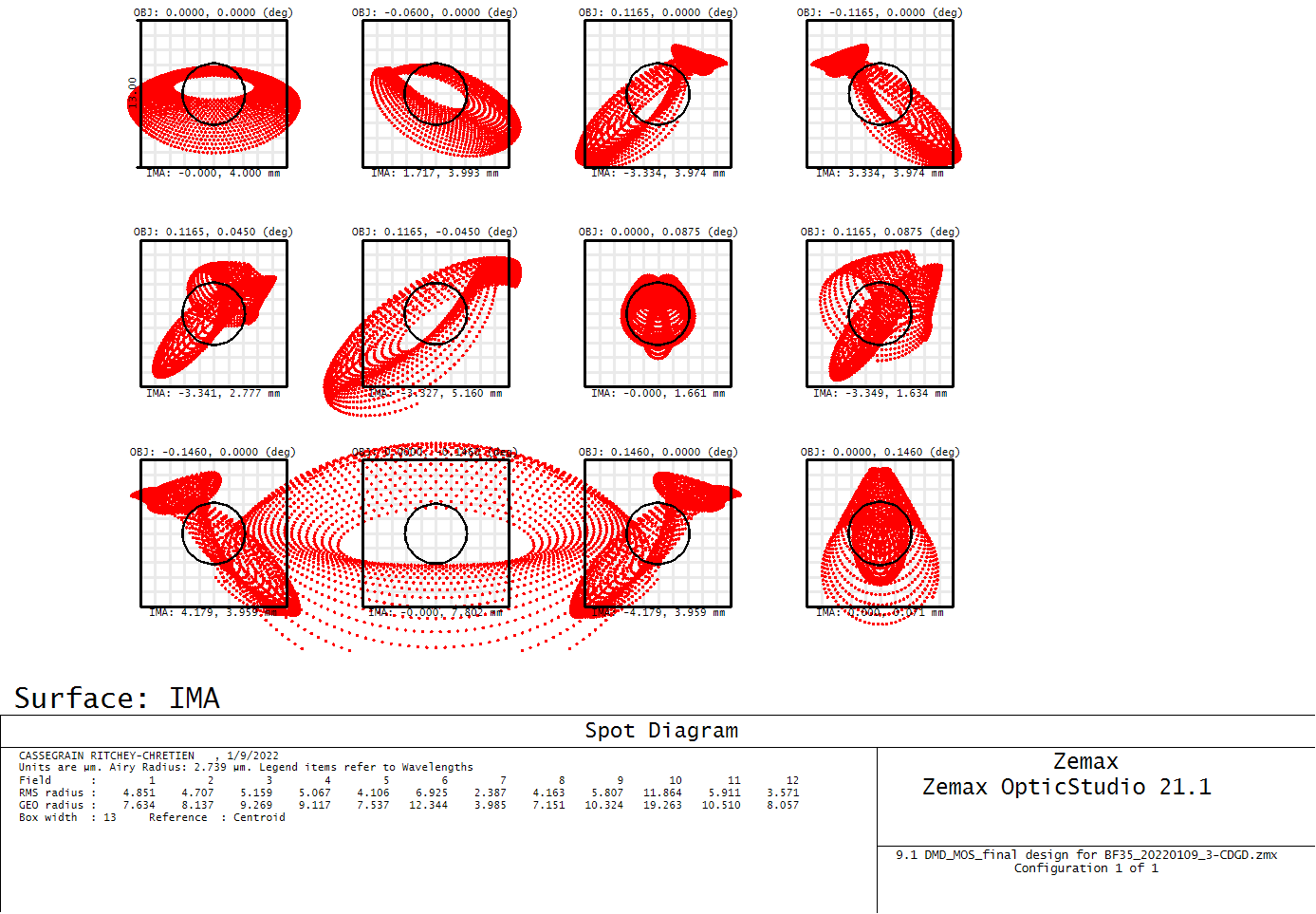}
        \caption{Spot diagrams at 700 nm }
    \end{subfigure}
    \begin{subfigure}{0.59\textwidth}
        \centering
        \includegraphics[width=0.9\textwidth]{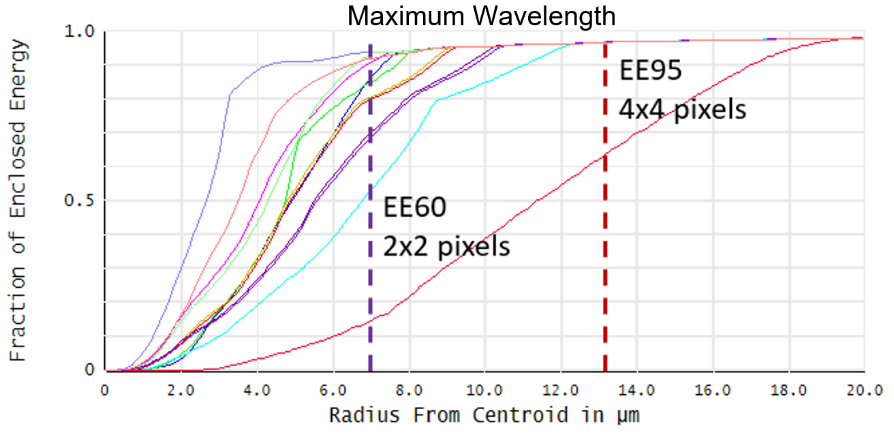}
        \caption{Enclosed energy curve at 700 nm}
    \end{subfigure}
    \caption{Performance for 700 nm; in the spot diagrams, each square is $2 \times 2$ pixels}
    \label{fig:700nm}
\end{figure}

\begin{figure}[H]
    \centering
    \includegraphics[width=0.7\textwidth]{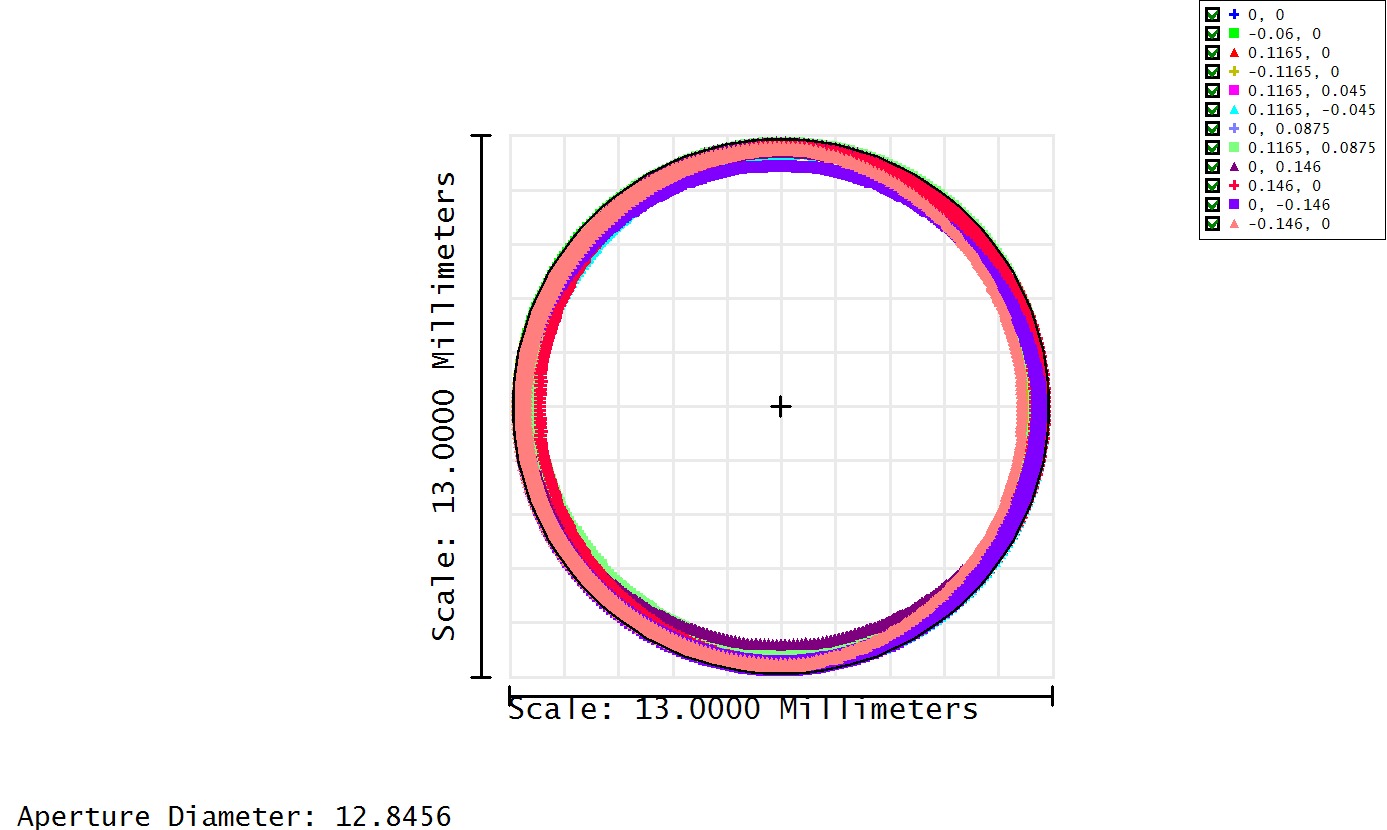}
    \caption{Footprint on pupil stop}
    \label{fig:pupil}
\end{figure}

\begin{figure}[H]
    \centering
    \includegraphics[width=0.85\textwidth]{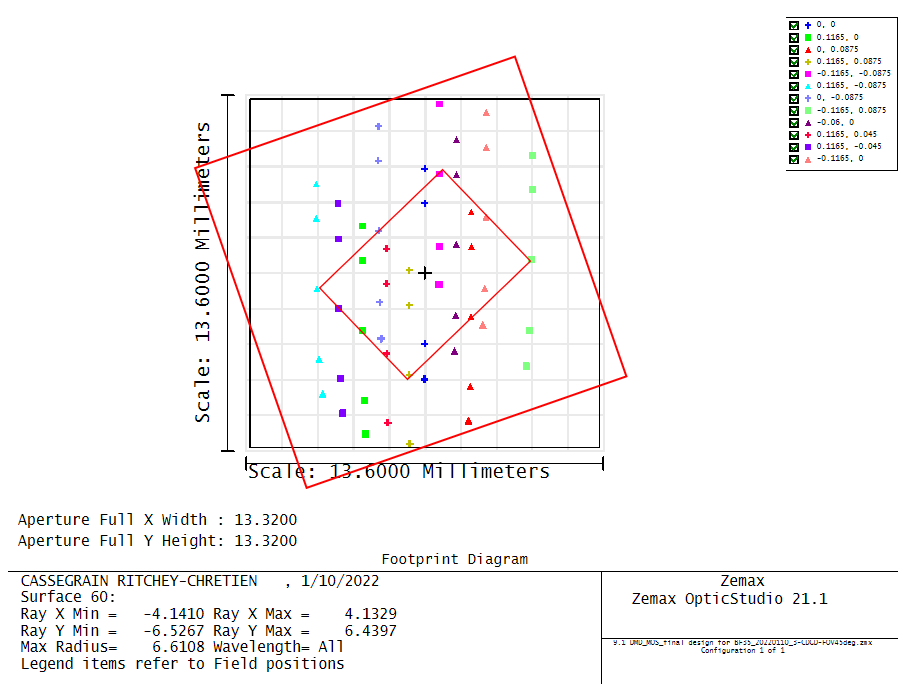}
    \caption{Footprint of spectra on image plane; color means different fields, and the wavelength range is from 400 nm to 700 nm. The outer and inner rectangular red outlines are those of the detector and DMD respectively.}
    \label{fig:footprint_spectra}
\end{figure}

A good quality pupil is generated just before the grating, which can benefit the consistent wavefront over the FOV and also provide an effective pupil stop. The footprint of the pupil is shown in Figure \ref{fig:pupil}. The footprint of the spectra is shown in Figure \ref{fig:footprint_spectra}, and the outlines of the DMD and detector are shown superimposed with a tilt of $45^\circ$.

\newpage
\subsubsection{Performance of Imaging Channel}

\begin{figure}[H]
    \centering
    \includegraphics[width=0.85\textwidth]{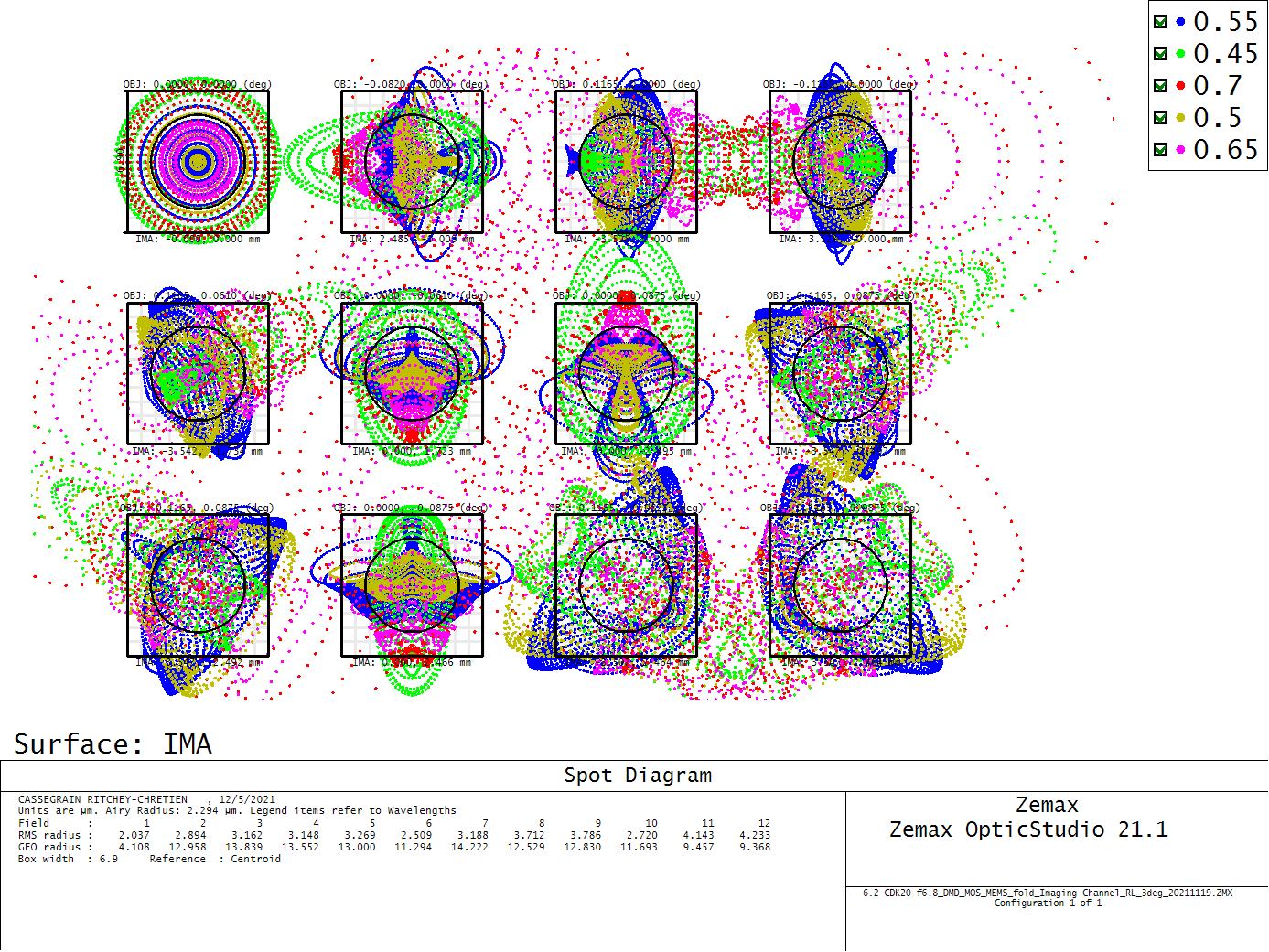}
    \caption{Spot diagrams of imaging channel}
    \label{fig:footprint_imaging_channel}
\end{figure}

The imaging channel is used for acquisition monitoring purposes, providing a parallel imaging function to monitor the full FOV. The optics relay the image on the DMD onto a CMOS detector. The detector in the imaging channel is a XIMEA MC031MG. Like the design of spectrograph, the optical performance in the wavelength range 400 to 450 nm is affected by the telescope optics, and so the optimization of the imaging channel was focused on 450 to 700 nm. Spot diagrams of the imaging channel are shown in Figure \ref{fig:footprint_imaging_channel}. 

\subsubsection{Transmission, Stray Light, and Sensitivity Analyses}

In addition to an analysis of image quality, analyses of transmission, stray light, and sensitivity have been conducted. The throughput of the spectrograph, including detector, is better than 30\%, and the maximum throughput is 48\% at center wavelength. For stray light analysis, a simulation was conducted with three sources, representing the center and edges of the DMD. In this simulation, a silver coating was applied for the mirrors FM1 and FM2, and an ideal coating with 99\% transmission was applied for the front and back surfaces of lenses. In addition, 80\% and 20\% efficiency were assigned for the 1st and 0th orders of the grating respectively, and lens barrels were applied for collimator and the camera lens groups. The result of this simulation showed that the ratio between stray light and signal on the detector was on the order of $10^{-5}$. A sensitivity analysis identified the most sensitive elements and allocated reasonable tolerances for manufacturing and alignment. 

\subsubsection{Status of Optical Design}

We have finalized the design of both channels of the DMD-MOS. For now, the design has been delivered to the vendor for manufacturing. A detailed assembly and testing plan is currently under development. 

\newpage
\section{DMD-MOS Wavelength Calibration}

\subsection{Introduction}

In this section, we will discuss a preliminary procedure to calibrate the spectrograph channel of the DMD-MOS. This procedure will be demonstrated on simulated ray tracing data from an older design of the DMD-MOS, due to time constraints at the time of writing. The main differences between our old design and current design is that the current design has improved optics, an improved VPH grism design, a smaller detector, and different rotational orientations between the DMD and detectors. Nevertheless, this procedure will inform a path forward for calibrating the DMD-MOS during assembly and testing. More details about this procedure can be found in another paper published in these proceedings \cite{Leung2022}.

Multi-object spectrographs, such as the DMD-MOS, suffer from hyperspectral imaging distortion. Due to grating aberrations, spectral lines will appear curved and parabola-shaped \cite{Schroeder2000}, rather than perfectly straight. The two types of hyperspectral imaging distortion, smile and keystone, need to be measured and corrected for accurate spectral and spatial calibration respectively\cite{Yokoya2010,Fisher1998}.


\subsection{Marked Slit Configuration}

\begin{figure}[H]
    \centering
    \begin{subfigure}{0.32\textwidth}
        \centering
        \includegraphics[width=0.75\textwidth]{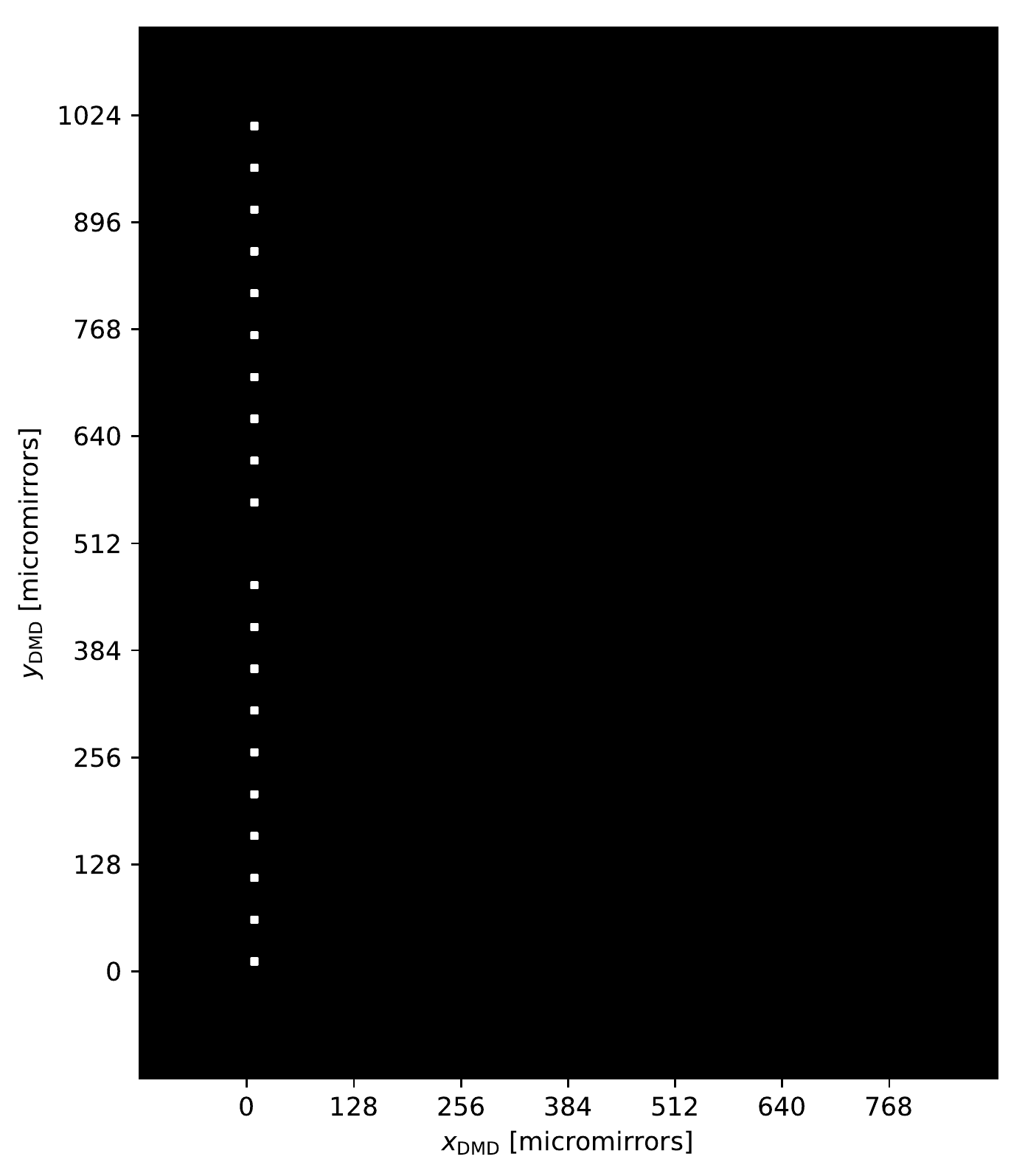}
        \caption{DMD with ON micromirrors along $x_\textrm{DMD}=9$}
        \label{fig:DMD-MOS_Footprint_DMD125}
    \end{subfigure}
    \begin{subfigure}{0.32\textwidth}
        \centering
        \includegraphics[width=0.75\textwidth]{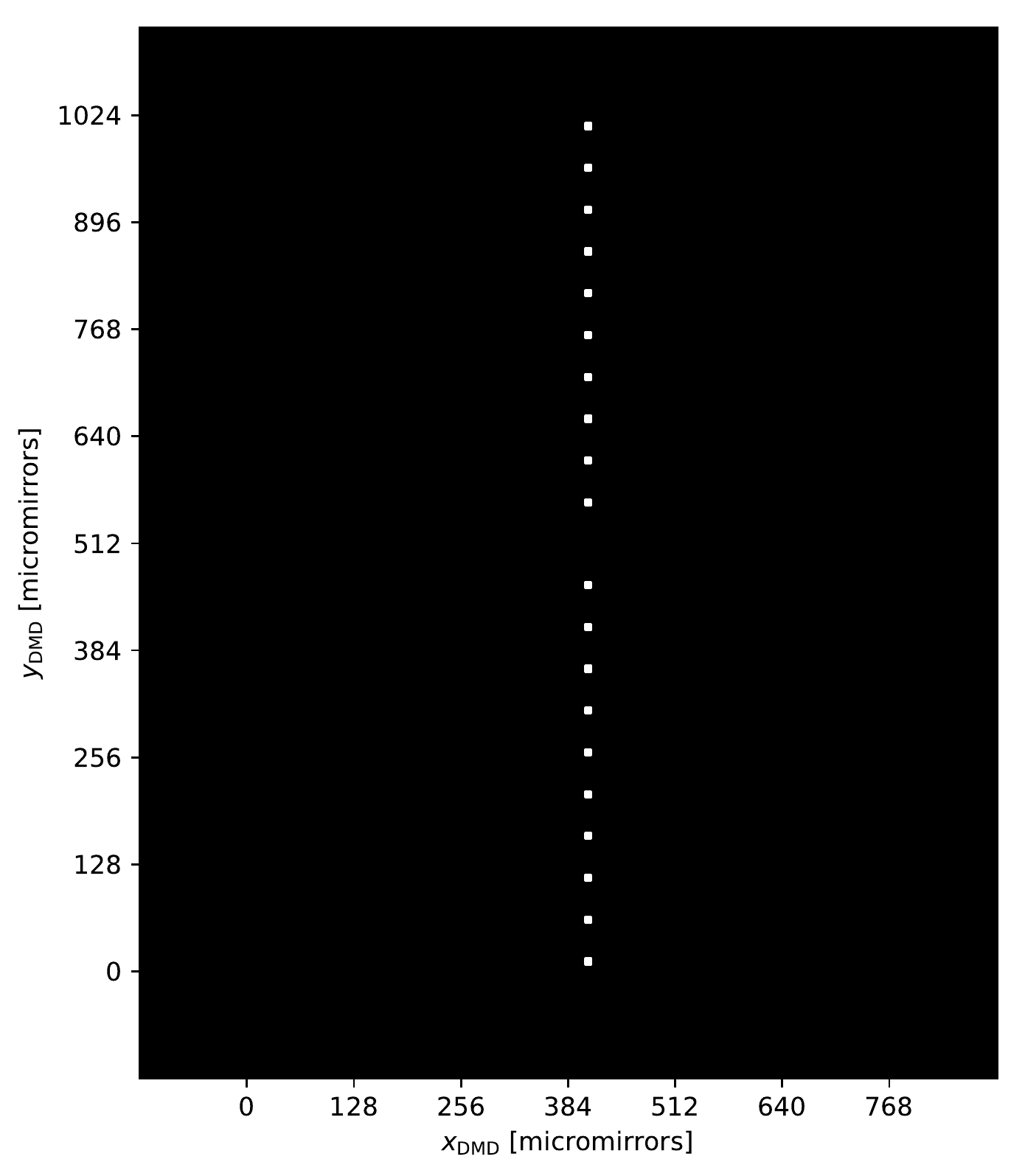}
        \caption{DMD with ON micromirrors along $x_\textrm{DMD}=408$}
        \label{fig:DMD-MOS_Footprint_DMD524}
    \end{subfigure}
    \begin{subfigure}{0.32\textwidth}
        \centering
        \includegraphics[width=0.75\textwidth]{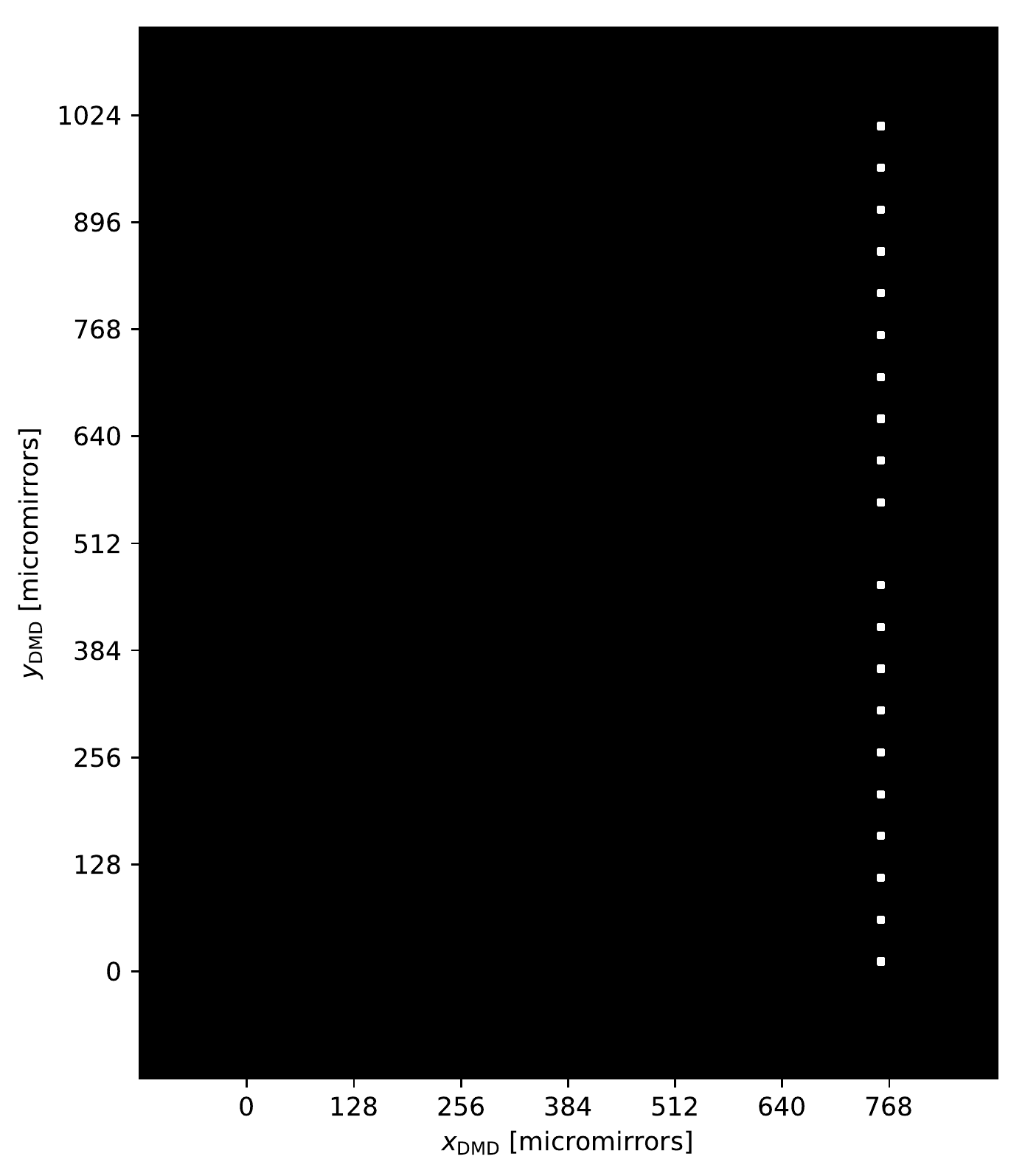}
        \caption{DMD with ON micromirrors along $x_\textrm{DMD}=758$}
        \label{fig:DMD-MOS_Footprint_DMD874}
    \end{subfigure}
    
    \begin{subfigure}{0.32\textwidth}
        \centering
        \includegraphics[width=\textwidth]{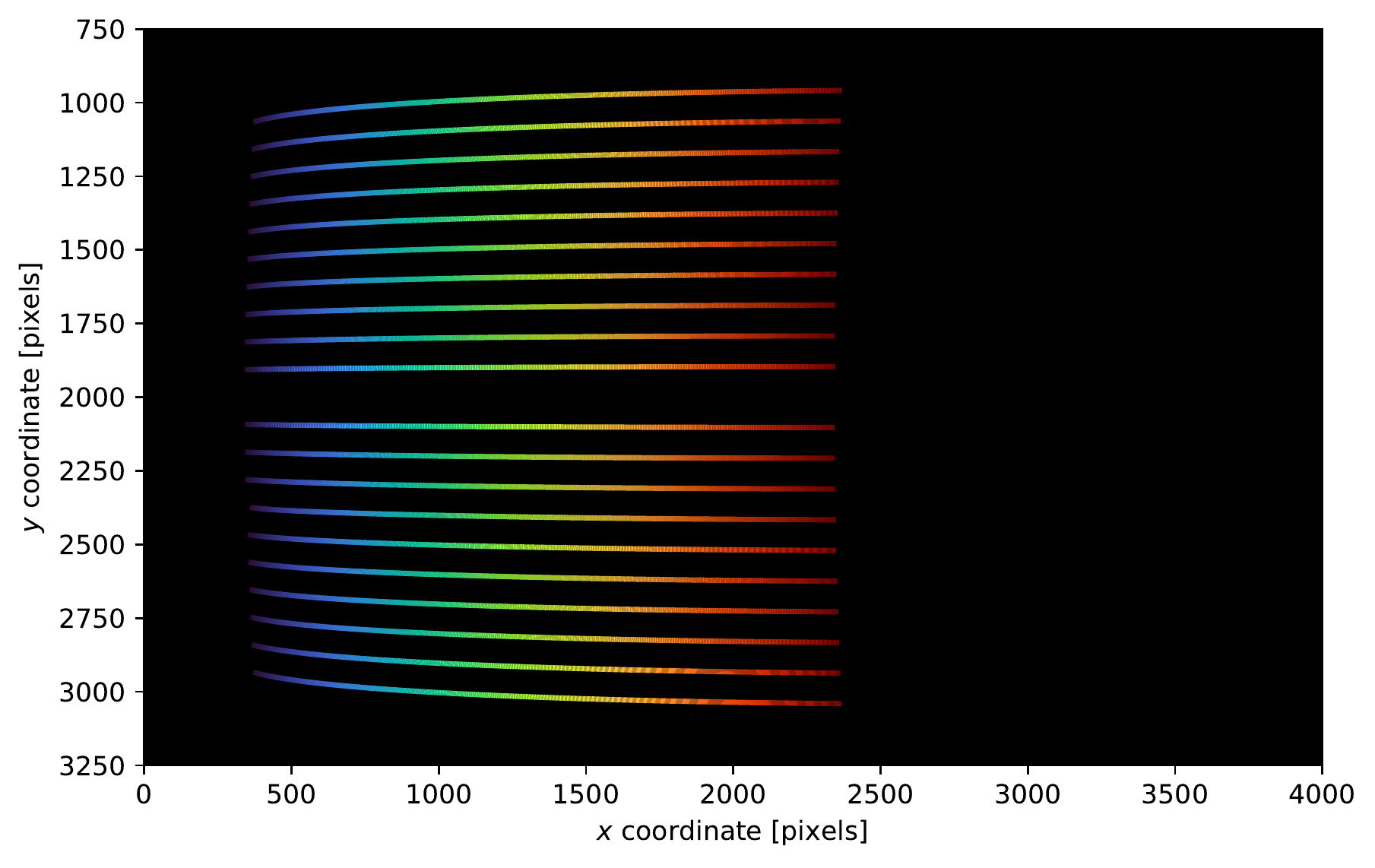}
        \caption{Corresponding spectra on detector for $x_\textrm{DMD}=9$}
        \label{fig:DMD-MOS_Footprint_Det125}
    \end{subfigure}
    \begin{subfigure}{0.32\textwidth}
        \centering
        \includegraphics[width=\textwidth]{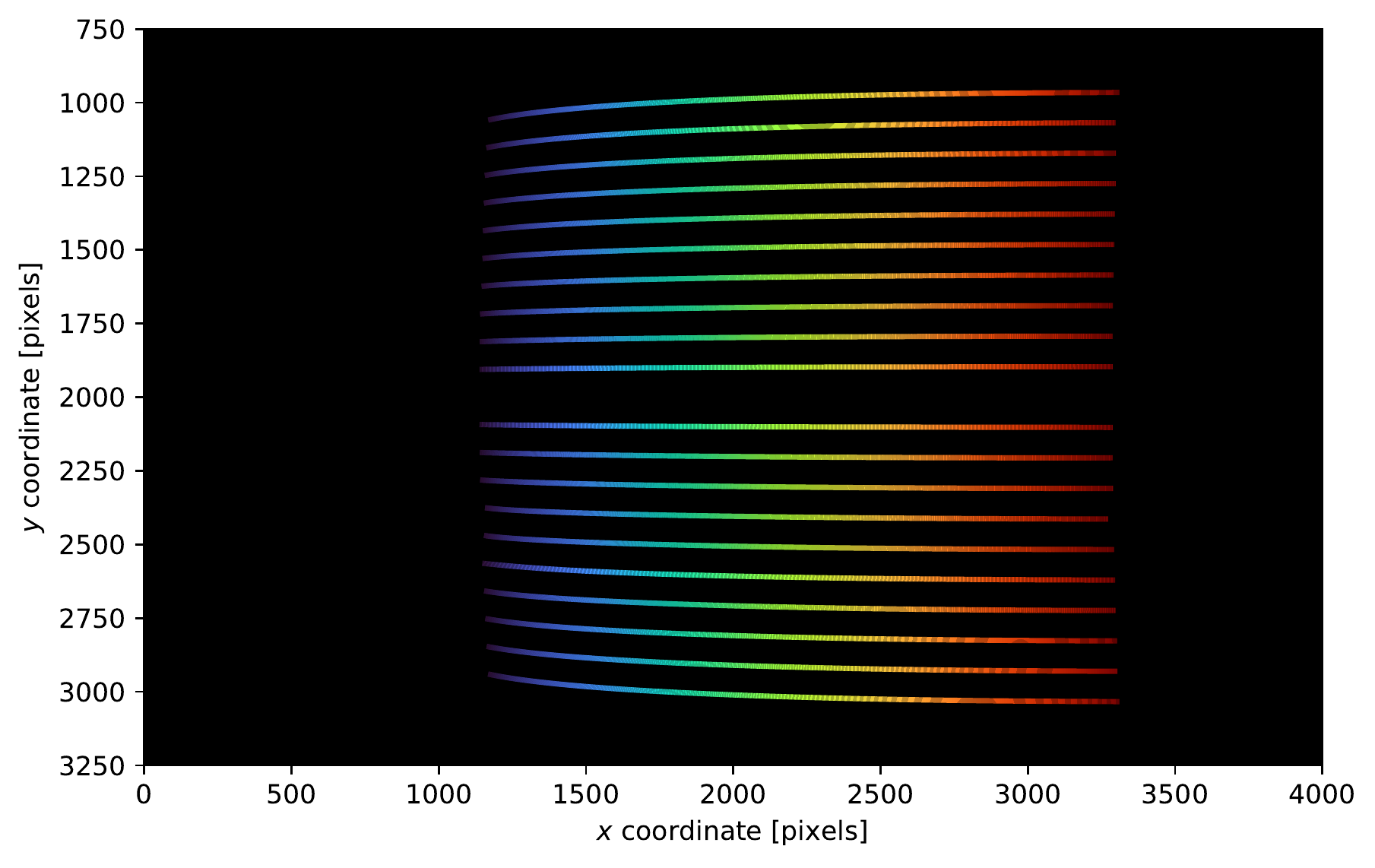}
        \caption{Corresponding spectra on detector for $x_\textrm{DMD}=408$}
        \label{fig:DMD-MOS_Footprint_Det524}
    \end{subfigure}
    \begin{subfigure}{0.32\textwidth}
        \centering
        \includegraphics[width=\textwidth]{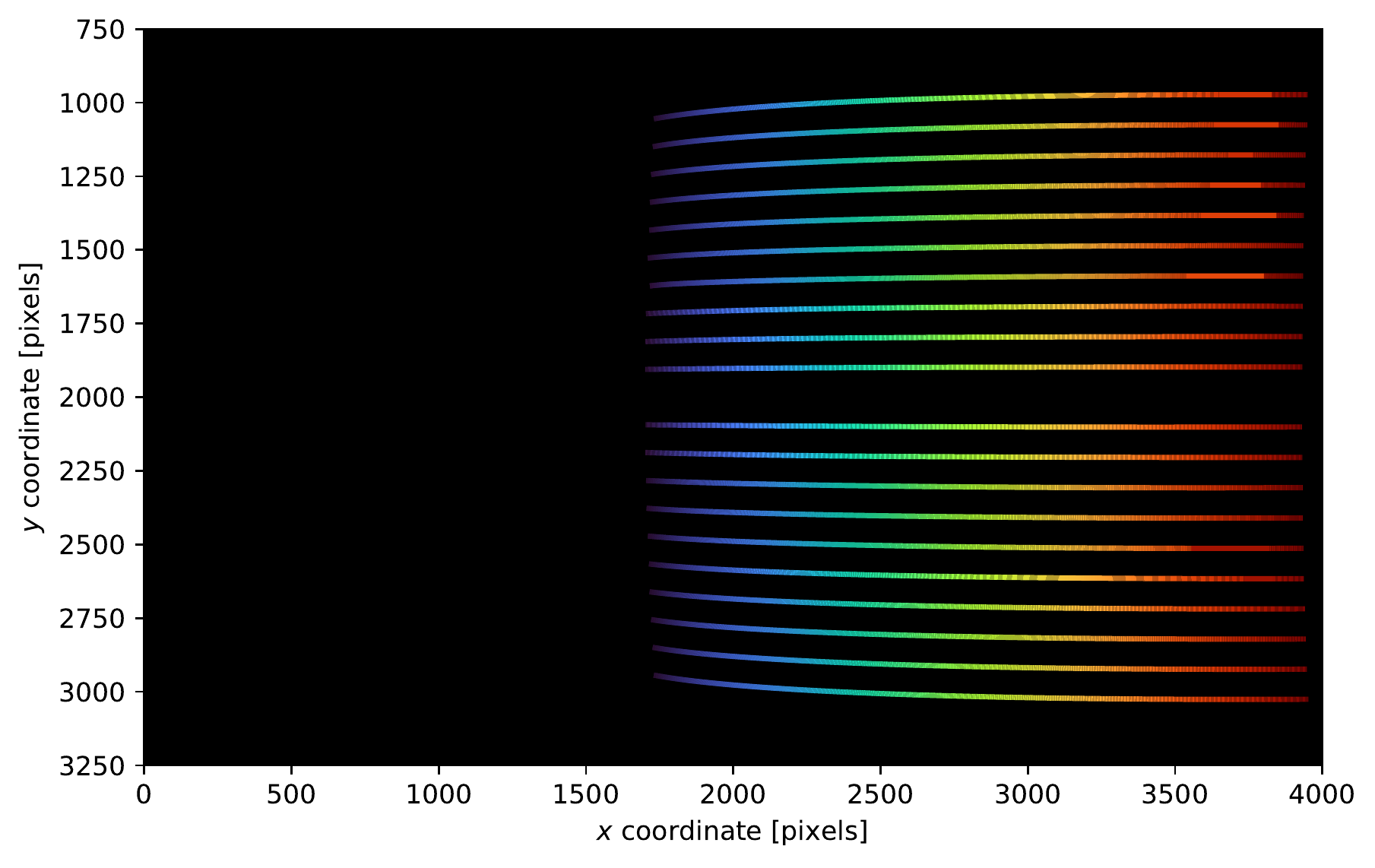}
        \caption{Corresponding spectra on detector for $x_\textrm{DMD}=758$}
        \label{fig:DMD-MOS_Footprint_Det874}
    \end{subfigure}
    \caption{Examples of some marked slit configurations for the DMD micromirrors (top row), and simulated data for the resulting spectra on the detector (bottom row). The ON micromirrors are shown as white squares in the top row figures, and have been enlarged for visibility.}
    \label{fig:DMD-MOS_Footprint}
\end{figure}

In order to calibrate the DMD-MOS, the micromirrors should be programmed to have a marked slit configuration. A marked slit has periodic holes in order to separate the spatial fields, allowing for keystone measurement\cite{Hong17}. With the DMD acting as a programmable slit, a marked slit configuration can be achieved by programming micromirrors to be in a spatially periodic ON configuration. Examples of some of these micromirror configurations and the resulting simulated spectra are shown in Figure \ref{fig:DMD-MOS_Footprint}. ON micromirrors are along a single line. By acquiring spectra for various micromirror configurations (various positions for the line of ON micromirrors), a general relationship between the DMD micromirrors, wavelength, and detector pixels could be developed, and this is the end goal.


\subsection{Smile and Keystone Fits}

\begin{figure}[H]
    \centering
    \begin{subfigure}{\textwidth}
        \centering
        \includegraphics[width=0.395\textwidth]{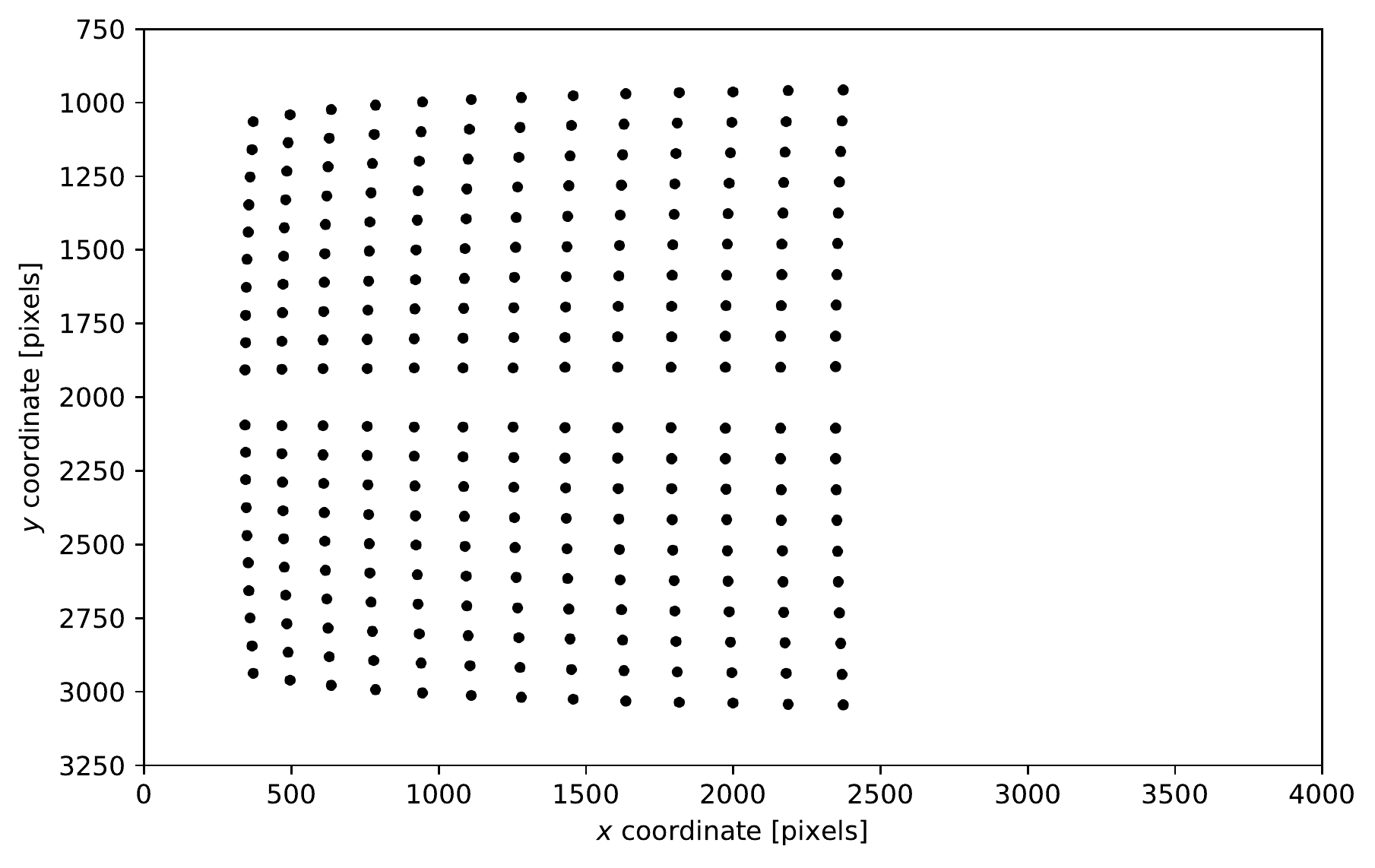}
        \caption{Simulated centroid points for the DMD micromirror configuration in Figure \ref{fig:DMD-MOS_Footprint_DMD125}}
        \label{fig:DMD-MOS_125_S}
    \end{subfigure}
    
    \begin{subfigure}{0.49\textwidth}
        \centering
        \includegraphics[width=0.808\textwidth]{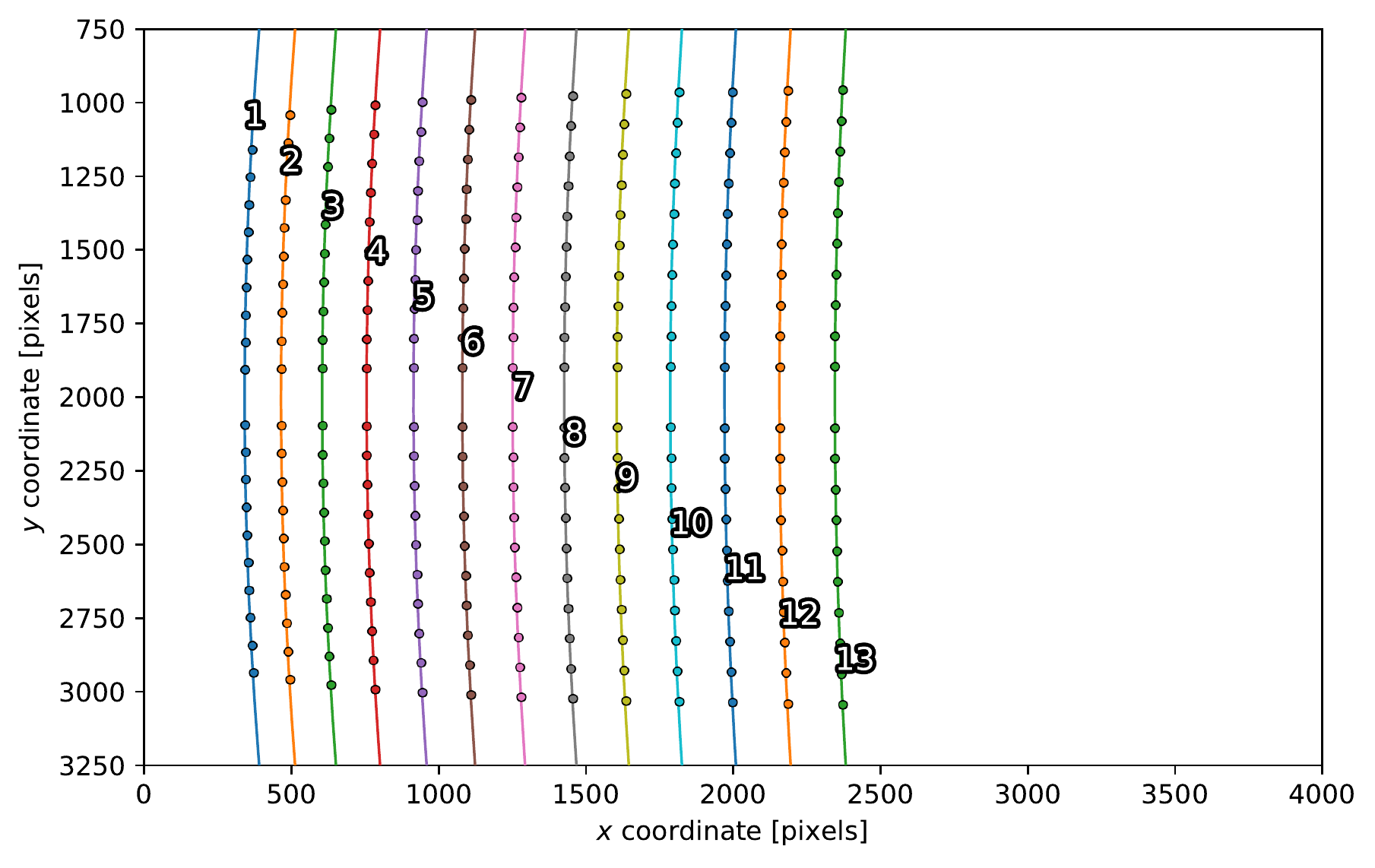}
        \caption{Smile fits to points in Figure \ref{fig:DMD-MOS_125_S}}
        \label{fig:DMD-MOS_125_clusters}
    \end{subfigure}
    \begin{subfigure}{0.49\textwidth}
        \centering
        \includegraphics[width=0.808\textwidth]{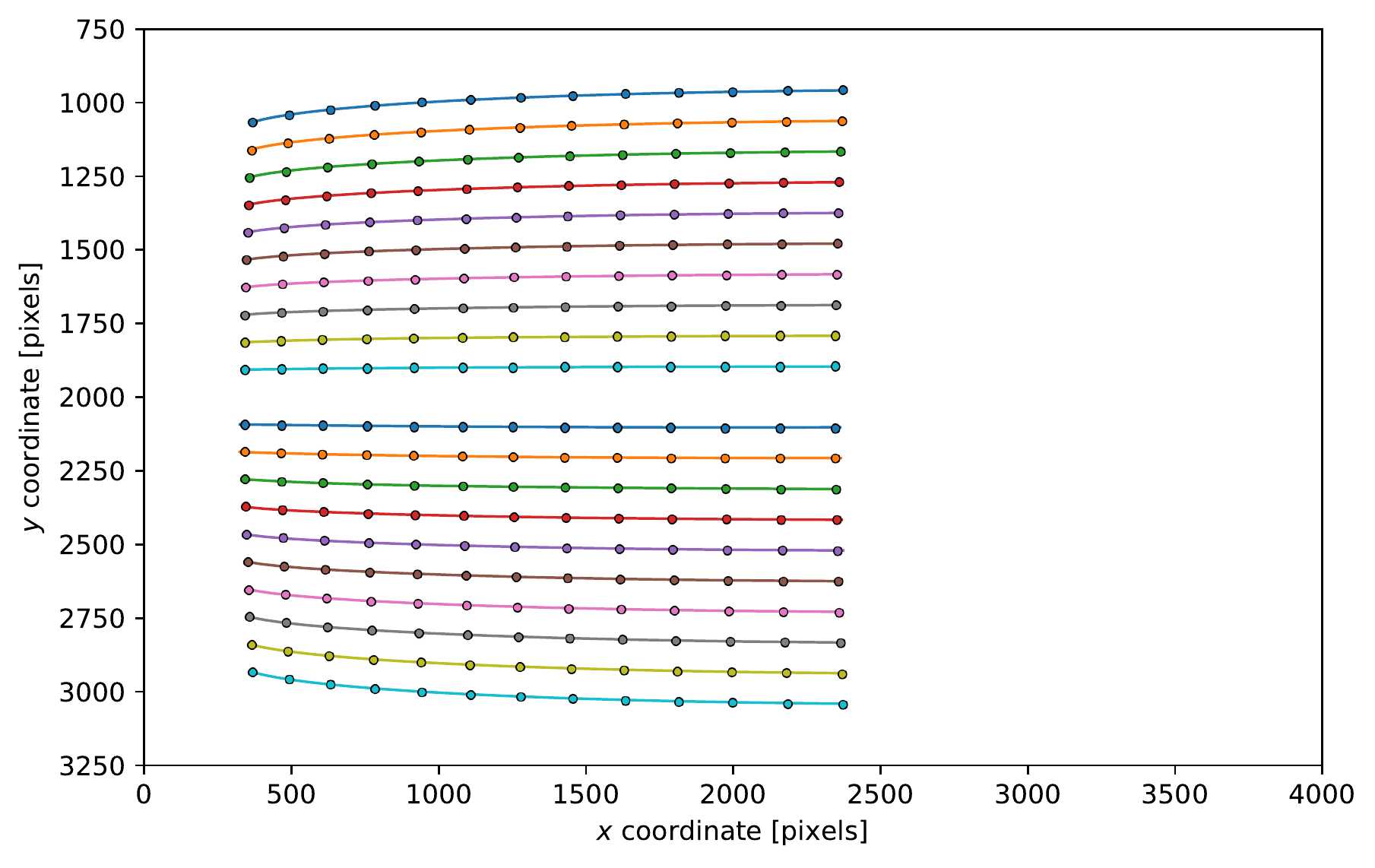}
        \caption{Keystone fits to points in Figure \ref{fig:DMD-MOS_125_S}}
    \end{subfigure}
    \caption{Smile and keystone fits for simulated DMD-MOS data corresponding to the DMD configuration in Figure \ref{fig:DMD-MOS_Footprint_DMD125}}
    \label{fig:DMD-MOS_125}
\end{figure}

In the ray tracing simulations that were conducted, for each of the micromirror configurations, $\sim$10000 rays were launched into the system at 13 different wavelengths ranging from 400 nm to 700 nm. Rays on the detector that were from the same micromirror and of the same wavelength were then grouped together, and their centroid coordinate was computed. An example of these centroid points are shown in Figure \ref{fig:DMD-MOS_125_S}, for the micromirror configuration in Figure \ref{fig:DMD-MOS_Footprint_DMD125}. If we were to calibrate the DMD-MOS in real life, the centroid points would be the locations of spectral peaks. For each wavelength value, a parabola can be fitted to the points corresponding to that wavelength. With the equations of these parabolas, smile can be measured across the detector. 

\begin{figure}[H]
    \centering
    \begin{subfigure}{0.49\textwidth}
        \centering
        \includegraphics[width=0.94\textwidth]{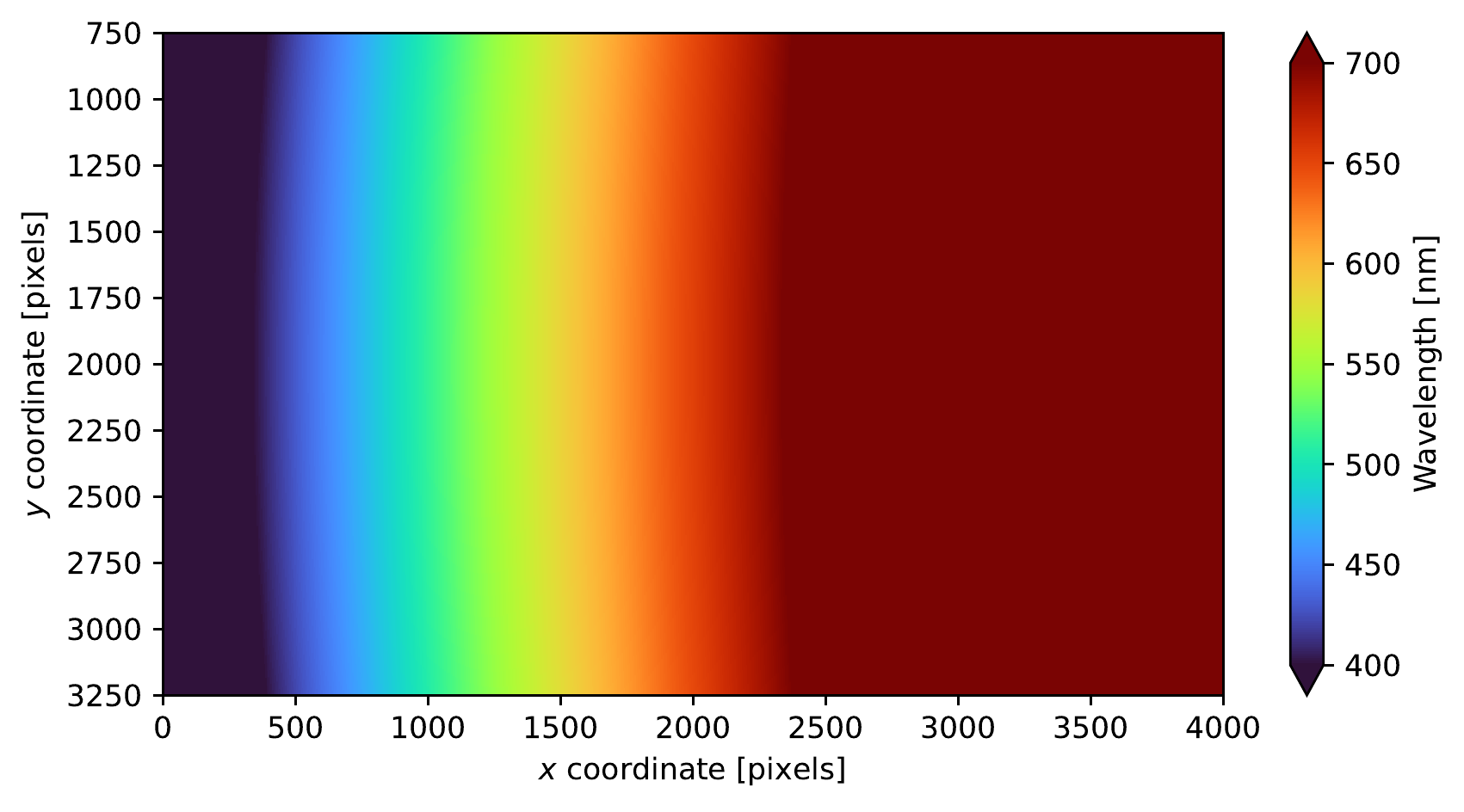}
        \caption{Pixel-to-wavelength mapping}
    \end{subfigure}
    \begin{subfigure}{0.49\textwidth}
        \centering
        \includegraphics[width=0.94\textwidth]{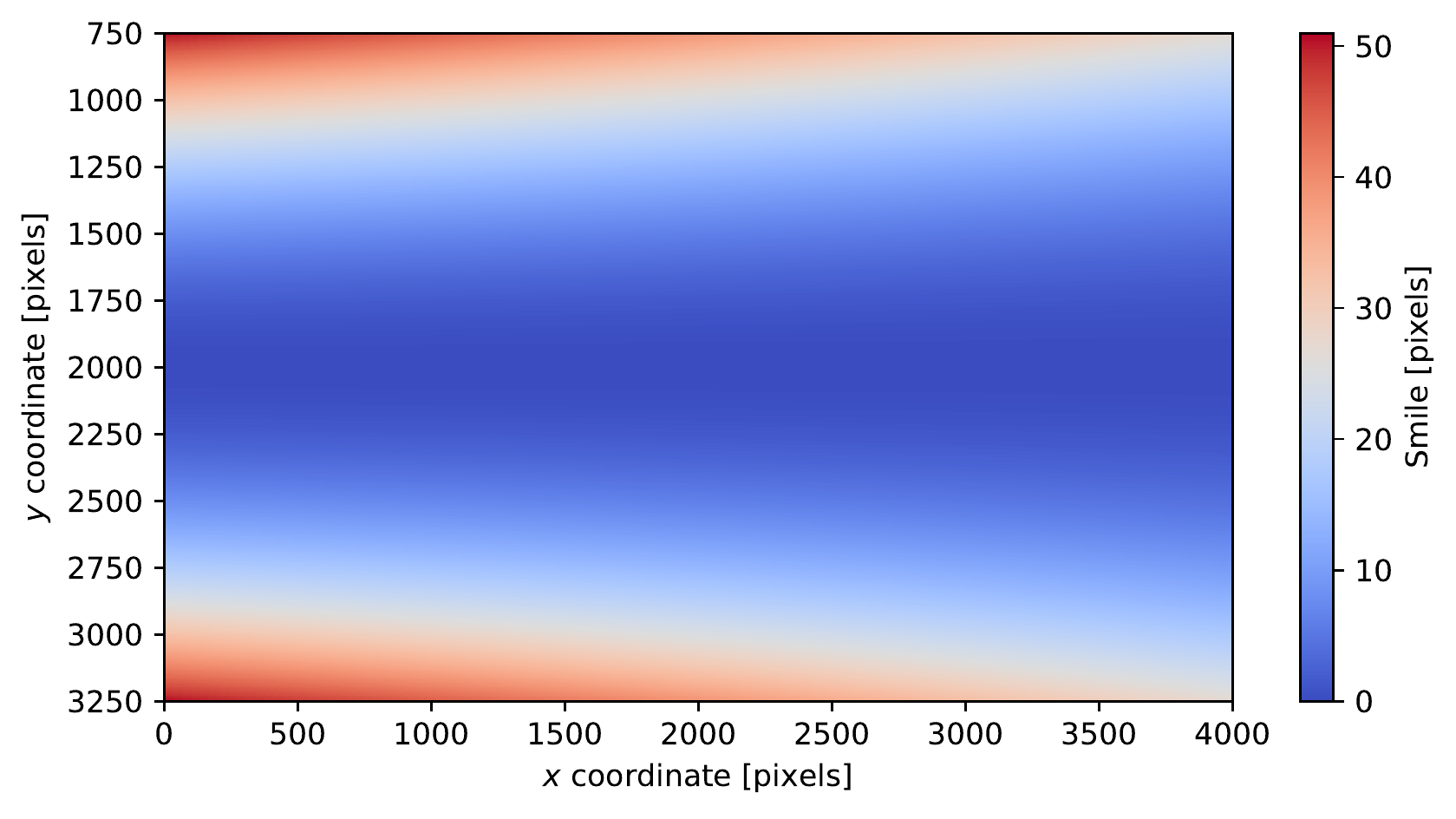}
        \caption{Pixel-to-smile mapping}
    \end{subfigure}
    \caption{Pixel-to-wavelength mapping and pixel-to-smile mapping for the simulated data in Figure \ref{fig:DMD-MOS_125_S}}
    \label{fig:DMD-MOS_125_w_s}
\end{figure}

For each ON micromirror, an ellipse can be fitted to the points corresponding to that micromirror. With the equations of these ellipses, keystone can be measured across the detector. 
With the equations of the parabola representing each spectral line (each wavelength value), a pixel-to-wavelength mapping and a pixel-to-smile mapping can be constructed. The parabolas in Figure \ref{fig:DMD-MOS_125_clusters} have the form $x=a(y-h)^2+k$ where $a$, $h$, and $k$ are some parameters. By modelling the relationships between these parameters and wavelength as polynomials, mappings can be constructed. We found that $a$ can be represented as a linear function of $k$, and wavelength can be represented as a cubic function of $k$. The resulting mappings for the simulated data in Figure \ref{fig:DMD-MOS_125_S} are shown in Figure \ref{fig:DMD-MOS_125_w_s}.

\subsection{Towards a General Relationship}

So far, relationships have been developed between the parabola parameters from the smile fits and the spectral coordinate. Analogous relationships need to be developed for keystone in order to relate the ellipse parameters from the keystone fits with the spatial coordinate, and this would be a next step. With the relationships between the parabola and ellipse parameters and the pixel-to-wavelength mappings for each micromirror configuration, the next step would be to use the data and mappings to develop a general relationship between DMD micromirror position, detector pixel position, and wavelength. 

\begin{figure}[H]
    \centering
    \begin{subfigure}{0.49\textwidth}
        \centering
        \includegraphics[width=\textwidth]{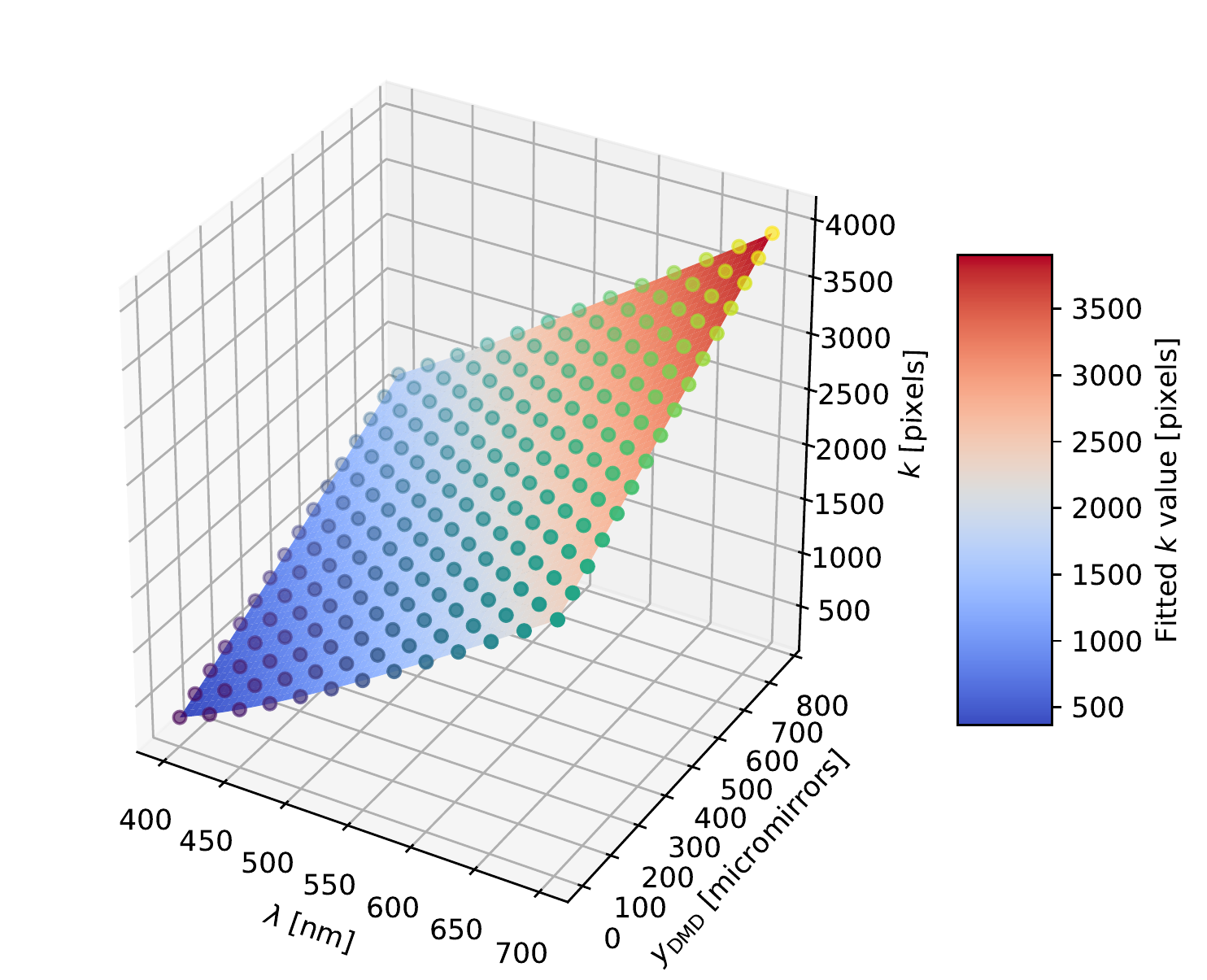}
        \caption{Fitted model}
    \end{subfigure}
    \begin{subfigure}{0.49\textwidth}
        \centering
        \includegraphics[width=\textwidth]{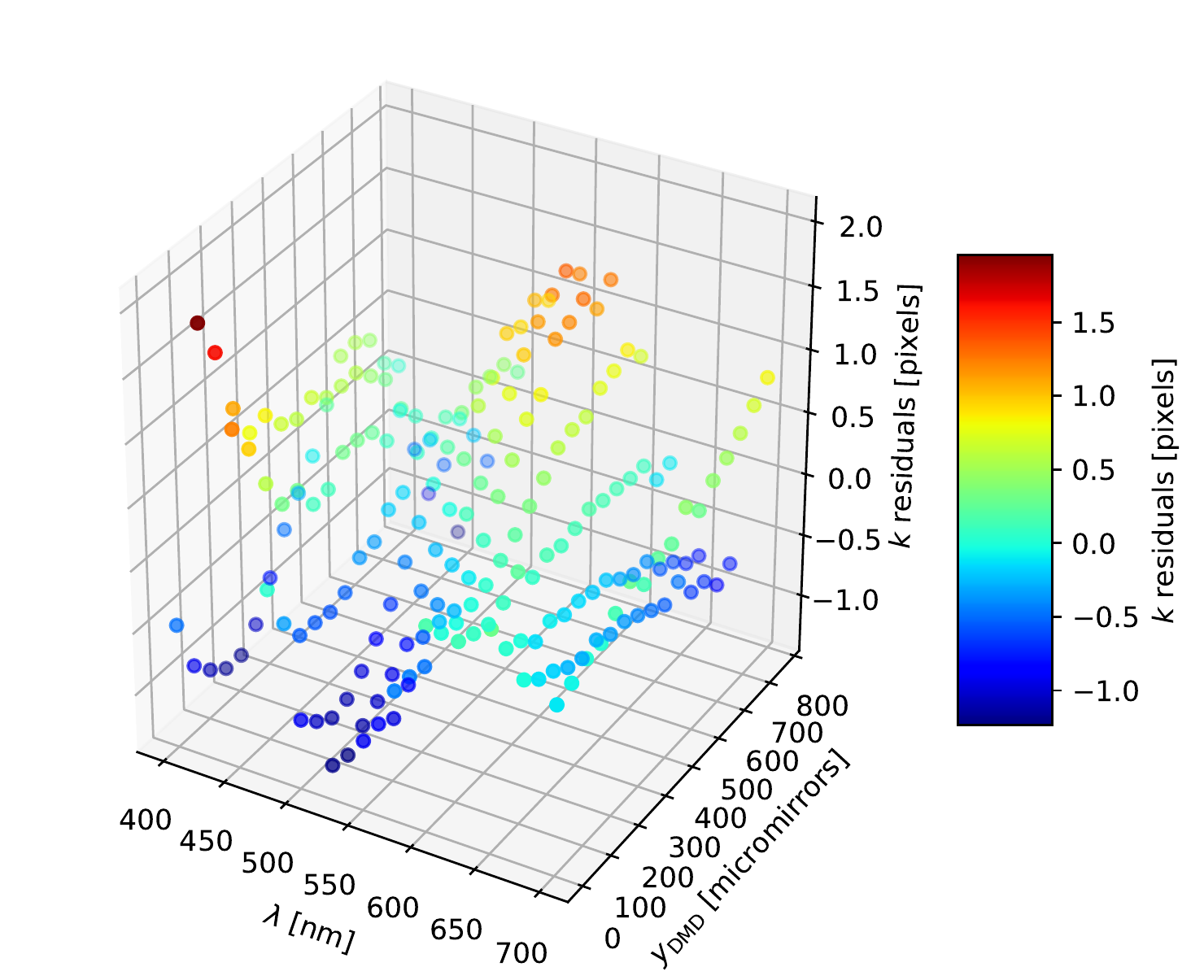}
        \caption{Residuals of model fit}
    \end{subfigure}
    \caption{Preliminary model for $k$ as a function of wavelength $\lambda$ and DMD vertical micromirror position $y_\textrm{DMD}$}
    \label{fig:gen_rel}
\end{figure}

We conducted some preliminary investigations to develop models between DMD micromirror position, wavelength, and the parabola parameters. In one of these preliminary investigations, we tried to model $k$ as a function of wavelength $\lambda$ and DMD vertical micromirror position $y_\textrm{DMD}$. By rearranging the grating equation and applying geometric ray trace equations from the DMD to the detector, we obtained an equation for $k$ as a function of $\lambda$ and $y_\textrm{DMD}$, which did not account for aberrations. We then added a polynomial correction term (polynomial function in $y_\textrm{DMD}$ and $\lambda$) to this equation in an attempt to account for the aberrations, and took the new equation with the correction term as our model for $k$. The resulting fitted model is shown in Figure \ref{fig:gen_rel}. The mean absolute error of the fit is $\sim$0.5 pixels, but there are some points with high error. More investigation is required, and this would be a next step.

\section{Summary}

DMDs are a promising technology for the programmable slit mask of MOS systems. We have designed a seeing-limited DMD-based MOS with a spectral resolution of $R\sim1000$ for a 0.5 m diameter F/6.8 telescope, as an exploratory study for future MOS systems. Our DMD-MOS covers a spectral range of 400 nm to 700 nm and has a FOV of $10.5^\prime \times 13.98^\prime$. The DMD-MOS contains two channels: a spectrograph channel to acquire spectra of objects, and an imaging channel for real-time monitoring of the slit configurations. These two channels offer parallel imaging and are designed around the DMD. Micromirrors in the DMD can be individually programmed to reflect light into either of the two channels, hence acting as slits in a conventional spectrograph. The dispersive element of this DMD-MOS is a VPH grism, which was used for its high throughput and to help minimize aberrations. The performance of our DMD-MOS was analyzed, and was found to have met the design requirements. In addition, the performance of the DMD was tested in lab, and the results were in agreement with those of other studies. Analyses of transmission and stray light were also conducted, as a well as a sensitivity analysis for tolerancing.

A preliminary procedure was developed to calibrate the spectrograph channel of the DMD-MOS, using simulated data. The DMD micromirrors were arranged in a marked slit configuration in order to separate the spatial fields. The resulting pattern on the detector was a set of points. For all the points corresponding to the same wavelength, a parabola was fitted to these points in order to measure smile distortion. From these parabolas, a pixel-to-wavelength mapping can be constructed for the detector. Keystone can be measured by fitting an ellipse to points corresponding to the same micromirror, but more investigation is needed. By analyzing spectra from various marked slit configurations, a general relationship between DMD micromirror position, detector pixel position, and wavelength could be developed. Some investigations were conducted for this general relationship, but more work is needed. The next steps are to continue developing this general relationship.

The design of the DMD-MOS is complete, and its parts are currently being manufactured by vendors at the time of writing. Assembly and testing of the DMD-MOS is expected to occur later this year. In the meantime, the electronics design of the DMD-MOS will be carried out, as well as further studies into the wavelength calibration procedure.

\acknowledgments 

The DMD-MOS project gratefully acknowledges the financial support provided by the Changchun Institute of Optics, Fine Mechanics and Physics (CIOMP) in the form of seed funding for this collaboration between the University of Toronto and CIOMP. The authors would like to thank PlaneWave Instruments for providing technical support. The authors would also like to thank Zemax for their support through the SMART Research Program. M.C.H. Leung would like to thank the Dunlap Institute for Astronomy and Astrophysics for their support of this research in the form of a Summer Undergraduate Research Program (SURP) Fellowship.
 
\bibliography{report} 
\bibliographystyle{spiejour} 

\end{document}